\documentclass[10pt,letterpaper]{article}

\usepackage[T1]{fontenc}
\usepackage[utf8]{inputenc}
\usepackage{newtxtext,newtxmath}
\usepackage{microtype}
\usepackage[letterpaper,top=0.68in,bottom=0.72in,left=0.74in,right=0.74in,
            headheight=15pt,headsep=16pt,footskip=26pt]{geometry}
\usepackage{graphicx}
\usepackage{float,needspace,placeins}
\usepackage{xcolor}
\usepackage{tikz}
\usetikzlibrary{arrows.meta,positioning,calc}
\usepackage{booktabs}
\usepackage{array}
\usepackage{tabularx, longtable}
\usepackage{enumitem}
\usepackage{amsmath}
\usepackage{titlesec}
\usepackage{fancyhdr}
\usepackage[font=small,labelfont={bf,rm},labelsep=period]{caption}
\usepackage[numbers,square,sort&compress]{natbib}
\usepackage[most]{tcolorbox}
\usepackage{xurl}
\usepackage{hyperref}
\definecolor{inkgreen}{HTML}{20584F}

\definecolor{Cobalt}{HTML}{1951E5}
\definecolor{Teal}{HTML}{39BCAE}
\definecolor{Ink}{HTML}{192238}
\definecolor{Subtle}{HTML}{637084}
\definecolor{SoftRule}{HTML}{DDE4EF}
\definecolor{Pale}{HTML}{F5F8FC}
\definecolor{PaleBlue}{HTML}{EDF3FF}
\hypersetup{colorlinks=true,linkcolor=Cobalt,citecolor=Cobalt,urlcolor=Cobalt,
  pdftitle={{Precision Atlas} for Scientific Kernels and Its Application to Adaptive Mesh Refinement},
  pdfauthor={Vansh Sharma and Venkat Raman}}

\newcommand{\papertitle}{\textit{Precision Atlas} for Scientific Kernels and Its \\ Application to Adaptive Mesh Refinement}
\newcommand{\paperauthors}{%
  Vansh Sharma\textsuperscript{1,2,*},
  Venkat Raman\textsuperscript{1,2}}
\newcommand{\paperdate}{November 2026}
\newcommand{\paperaffiliations}{%
  \textsuperscript{1}Center for Prediction, Reasoning \& Intelligence for Multiphysics Exploration (C-PRIME)\\[2pt]
  \textsuperscript{2}Department of Aerospace Engineering, University of Michigan}

\newcommand{\papercorrespondence}{Vansh Sharma (\href{mailto:vanshs@umich.edu}{vanshs@umich.edu})}

\color{Ink}
\setlist[itemize]{topsep=3pt,itemsep=3pt,parsep=0pt,leftmargin=1.35em,
  label={\textcolor{Cobalt}{\large\textbullet}}}
\titleformat{\section}{\rmfamily\bfseries\large\color{Ink}}
  {\textcolor{Cobalt}{\thesection}}{0.70em}{}
\titleformat{\subsection}{\rmfamily\bfseries\normalsize\color{Ink}}
  {\textcolor{Cobalt}{\thesubsection}}{0.70em}{}
\titleformat{\subsubsection}{\rmfamily\bfseries\normalsize\color{Ink}}
  {\textcolor{Cobalt}{\thesubsubsection}}{0.70em}{}
\titlespacing*{\subsubsection}{0pt}{8pt plus 1pt minus 1pt}{3pt}
\titlespacing*{\section}{0pt}{17pt plus 2pt minus 1pt}{6pt}
\titlespacing*{\subsection}{0pt}{11pt plus 1pt minus 1pt}{4pt}
\fancypagestyle{cprime}{%
  \fancyhf{}%
  \fancyhead[L]{\rmfamily\fontsize{8}{10}\selectfont\bfseries\color{Ink}C-PRIME}%
  \fancyhead[R]{\rmfamily\fontsize{7.5}
  {9}\selectfont\color{Subtle}Precision Atlas for Scientific Kernels}%
  \fancyfoot[R]{\rmfamily\fontsize{8.5}{10}\selectfont\bfseries\color{Cobalt}\thepage}%
}
\fancypagestyle{cprimefirst}{%
  \fancyhf{}%
  \fancyfoot[L]{\rmfamily\fontsize{7.6}{9.2}\selectfont\color{Subtle}\textcopyright\ 2026 All rights reserved by the Author and C-PRIME.}%
  \fancyfoot[R]{\rmfamily\fontsize{8.5}{10}\selectfont\bfseries\color{Cobalt}\thepage}%
}
\newtcolorbox{abstractpanel}{
  enhanced,breakable,boxrule=0pt,colback=Pale,
  borderline west={2.2pt}{0pt}{Cobalt},
  sharp corners,left=13pt,right=12pt,top=11pt,bottom=10pt,
  before skip=0pt,after skip=8pt
}
\newcommand{\minilabel}[1]{{\rmfamily\fontsize{8.5}{10}\selectfont\bfseries\color{Cobalt}\MakeUppercase{#1}}}

\begin{document}
\thispagestyle{cprimefirst}

\noindent\begin{minipage}[c]{0.60\linewidth}
  \includegraphics[width=1.72in]{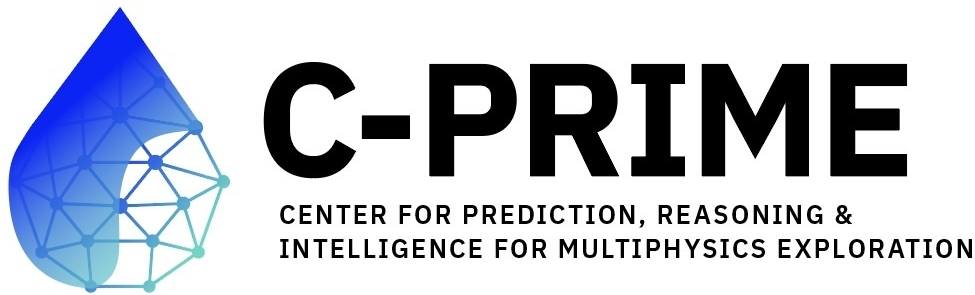}
\end{minipage}\hfill\begin{minipage}[c]{0.37\linewidth}
  \raggedleft\rmfamily\fontsize{7.6}{10.2}\selectfont
  \color{Subtle}RESEARCH SERIES\quad$\bullet$\quad \paperdate
\end{minipage}

\vspace{2pt}
\noindent\textcolor{Cobalt}{\rule{0.62in}{1.6pt}}%
\textcolor{SoftRule}{\rule{\dimexpr\linewidth-0.62in\relax}{0.8pt}}
\vspace{10pt}

{\rmfamily\bfseries\fontsize{23}{27}\selectfont\color{Ink}\papertitle\par}
\vspace{5pt}
{\rmfamily\fontsize{9.9}{12.4}\selectfont\bfseries\color{Ink}\paperauthors\par}
\vspace{4pt}
{\rmfamily\fontsize{8.9}{12.2}\selectfont\color{Subtle}\paperaffiliations\par}
\vspace{12pt}

\begin{abstractpanel}
\minilabel{Abstract}\par\vspace{3pt}

In scientific calculations, the precision of the stored quantities affects the accuracy of the solution and, in adaptive methods, the spatial resolution selected by the numerical method. The objective of this work is to determine the precision required by individual numerical operations and to apply these requirements to adaptive mesh refinement (AMR). For this purpose, selected Lean-verified arithmetic results, analytical precision bounds and 6,480 evaluations of 60 scientific kernels are organized in a ``Precision Atlas''. Controlled time integration problems are first used to distinguish the requirements for time advancement, state accumulation and output. The analysis is then extended to AMR, where a rounded sensor determines which cells are refined. The probabilities of tagging a cell and selecting a refinement block under independent stochastic rounding are derived, accounting for shared stencil values. It is shown that rounding can tag a uniform field, producing a phantom gradient, when the floating-point spacing exceeds the refinement threshold. For a fixed gradient threshold, excluding these tags requires one additional significand bit for each halving of the mesh spacing. Shock--vortex and Rayleigh--Taylor calculations confirm that the extra refinement appears where this condition is violated. In the finer shock--vortex case, stochastic bfloat16 (BF16) tagging requires 2.69 times the cell updates of double-precision (FP64) tagging. The Atlas bounds are then used to construct a tagging criterion that excludes these extra tags. This criterion retains 20.03\%, 61.67\% and 99.99\% of the reference tags under stochastic BF16, FP16 and FP32 sensor rounding, respectively. With BF16, the lost tags remove refinement around the vortex and the Rayleigh--Taylor interface. When the conserved state is also stored in FP16, the bound must include pressure reconstruction from conserved variables, and 94.73\% of the decisions remain unresolved under stochastic rounding. The Atlas indicates when the stored quantities cannot establish the requested refinement decision and is a step toward \textit{a priori} precision selection for scientific calculations.

\par\vspace{7pt}
{\color{SoftRule}\rule{\linewidth}{0.55pt}}\par\vspace{2pt}
{\rmfamily\fontsize{8.5}{11.5}\selectfont\color{Subtle}
\textbf{\textcolor{Ink}{Keywords}}\quad Floating-point arithmetic; adaptive mesh refinement; stochastic rounding; scientific kernels; precision bounds; phantom gradients\par
\vspace{3pt}
\textbf{\textcolor{Ink}{Corresponding author}}\quad\papercorrespondence\par}
\end{abstractpanel}

\section{Introduction}\label{sec:intro}

The computational effort of resolving the relevant physical scales is a major consideration in the simulation of reacting flows, turbulent mixing and compressible fluid motion. A numerical method must resolve changes in the state associated with processes that can occur on disparate spatial and temporal scales. In a compressible flow calculation, for example, the pressure force depends on differences between pressures in adjacent cells, and the pressure itself may be evaluated as a function of the difference between total and kinetic energy. The accuracy of these differences determines which features can be distinguished numerically. Although spatial and temporal discretization have received considerable attention in the design of computational methods, the representation of the underlying physical quantities imposes an additional constraint. Decreasing the cell size reduces the difference associated with a given smooth gradient, and decreasing the timestep reduces the increment to a stored state. A representation that resolves the physical quantities on a coarse grid can therefore become inadequate as the numerical resolution increases. Of particular interest here are adaptive methods, in which these quantities also determine the locations at which the resolution is increased.

The choice of floating-point representation in scientific computing is increasingly influenced by the architecture of modern accelerators. IEEE half (FP16), single (FP32) and double precision (FP64, IEEE binary64) have significand precisions of 11, 24 and 53 bits, respectively; bfloat16 (BF16) has 8 bits~\cite{ieee7542019,higham2002accuracy}. Mixed-precision methods developed for machine learning have demonstrated that storing some quantities in lower-precision formats while evaluating sensitive operations or maintaining accumulations in higher precision can reduce computational cost~\cite{micikevicius2018mixed,micikevicius2022fp8}. For instance, the NVIDIA H100 architecture provides substantially higher peak throughput for FP8 tensor operations than for FP64 operations~\cite{nvidia2022hopper}. The extent to which scientific algorithms can exploit this disparity depends on the structure of the calculations: stencils, reductions, nonlinear iterations and conditional operations have different arithmetic and memory-access patterns. In computational fluid dynamics (CFD), Witherden and Jameson~\cite{witherden2020impact} examined the impact of number representation on high-order implicit large-eddy simulations, while Brogi et al.~\cite{brogi2024openfoam} studied precision-dependent solution quality and convergence in OpenFOAM. Karp et al.~\cite{karp2026effects} examined lower precision in several scale-resolving turbulence solvers and identified quantities with differing precision sensitivity. These studies provide evidence for selective precision assignment. Its effectiveness depends on which flow quantities or statistical measures must be resolved, as well as on the numerical operations that produce them.

An approach to determining the required arithmetic precision is to compare the errors introduced by rounding with those arising from the discretization. Tamstorf et al.~\cite{Tamstorf2021} established how matrix storage quantization in elliptic multigrid can become a dominant contribution as the spatial resolution increases, motivating a progressive allocation of precision across solver stages and grid levels. Kohl et al.~\cite{kohl2024multigrid} extended the representation choices through block floating-point arithmetic. More recently, Mazumder et al.~\cite{Mazumder2026} used a signal-to-noise criterion based on the discretization error to identify subcomputations for which the arithmetic precision can be reduced. The benefits of these approaches derive from the observation that a reduction in arithmetic error below the discretization error may have little effect on the desired result. A related issue is the representation of the physical signal itself. Two distinct pressures that encode to the same stored word cannot yield their exact difference from those words alone. When an unrounded fluctuation is available, a centered representation may allow this difference to be resolved using fewer bits. The lower bound imposed by indistinguishable inputs is familiar from information-based complexity~\cite{Traub1988}, and the sensitivity of a linear solve to the direction of the right-hand side is described by effective conditioning~\cite{ChanFoulser1988}. Such arguments establish conditions on the represented quantities; the error of the subsequent arithmetic must be assessed separately.

Similar considerations apply to the temporal resolution of physical phenomena. The upper bound on the timestep for an explicit method is determined by its stability properties, whereas the representable differences between simulation times and between stored states are determined by the corresponding formats. Shampine~\cite{Shampine1974} examined limits on useful steps and distinguishable time arguments in differential equation solvers. Of particular relevance to low-precision integration is the accumulation of state changes that are smaller than the spacing of the stored state. Croci and Giles~\cite{CrociGiles2023} analyzed this problem for the heat equation, demonstrating how stagnation under round-to-nearest arithmetic affects the time integration. Barad~\cite{Barad2026} separated the error due to state storage from that introduced during right-hand-side evaluation for Runge--Kutta and multistep integration; the state-storage term increases with decreasing timestep and can be reduced using a two-word representation. These results motivate the distinction between the precision of an incoming increment, the precision of the accumulator and the precision of the output used in subsequent operations. Maintaining a higher-precision global time variable is inexpensive in most solvers. The compatibility between time spacing and the timestep limit is relevant when a common working type is imposed or when many local time histories are represented.

Stochastic rounding (SR) provides a mechanism for retaining small changes in expectation. A value between two representable numbers is mapped to the upper number with probability proportional to its distance from the lower number~\cite{Connolly2021,CrociEtAl2022}. Such an encoding has zero conditional mean error, and the statistical accumulation of the errors can be favorable compared with repeated round-to-nearest (RN) operations. The reduced-precision climate simulations of Paxton et al.~\cite{paxton2022climate} and Kimpson et al.~\cite{kimpson2023climate} illustrate the utility of this approach. In their discussion of extending the coarse-resolution SPEEDY experiments to finer grids, Kimpson et al. identify the potential loss of small derivatives at reduced precision and propose scaling the values or evaluating the derivatives in higher precision before stochastic conversion for data transport (Section~4 of Ref.~\cite{kimpson2023climate}). This is important because a derivative acts on a difference of two represented quantities. Even if the signed difference of independently rounded uniform values has zero expectation, its absolute value can have positive expectation. In particular, an absolute difference used as a refinement indicator responds to the amplitude of the rounding fluctuations. We refer to such a rounding-induced difference in an otherwise uniform field as a \emph{phantom gradient}. The distinction between the moments of a rounded quantity and those of a nonlinear function of that quantity is familiar from dithered quantization~\cite{GrayStockham1993}. The question considered in this work is how this effect propagates through a spatial adaptation algorithm.

Adaptive mesh refinement (AMR) introduces a feedback between a numerical solution and its discretization. In block-structured AMR, as developed for shock hydrodynamics by Berger and Colella~\cite{BergerColella1989}, the cells or regions requiring higher resolution are determined from the numerical solution and the refined hierarchy is used in subsequent flux calculations. RAPTOR~\cite{RAPTOR2025} has demonstrated the practical utility of precision profiling and assignment by AMR level in Flash-X multiphysics simulations. A complementary aspect of precision allocation is the sensitivity of the hierarchy to the arithmetic of the refinement decision itself. In a gradient-based method, a refinement threshold divides local field differences into two classes; a rounding error near this threshold can switch the selected class. Unstable floating-point tests and the divergence between real-arithmetic and floating-point branches have been investigated in program analysis~\cite{DarulovaKuncak2014,Titolo2019}, and adaptive-precision filters use error estimates to determine when a predicate must be evaluated more accurately~\cite{Shewchuk1997}. These principles apply to individual AMR decisions. A particular feature of the spatial problem is that neighboring decisions depend on overlapping stencils, and a refinement block can be selected by any one of several buffered tags. The probability of refining the block consequently depends on a joint distribution of the rounded sensor values. Evaluating this distribution provides a means to relate precision to mesh size and computational work in addition to errors in the physical solution.

The objective of this study is to determine which physical variations a numerical representation can resolve and how the resulting limits affect adaptive calculations. Spatial refinement reduces the difference across a smooth field and, for explicit methods, can reduce the permitted timestep. Both effects change the precision required to represent the numerical signal. An operation-level precision requirement must therefore account for its role in both the evolution of the state and selection of the mesh.

\subsection{Research questions and contributions}\label{sec:rqs}

\begin{itemize}
\item \textbf{RQ1: Representation requirements.} How do physical scales, mesh spacing and the permitted timestep determine the precision required to represent differences, state increments and outputs?
\item \textbf{RQ2: Adaptive refinement.} How do the representation and rounding policy of a refinement sensor change local tags, the selected mesh and the physical solution?
\item \textbf{RQ3: Use of precision bounds.} How can an operation-specific reference of analytical bounds guide precision selection and distinguish rounding-induced refinement from resolvable physical variation?
\end{itemize}

The contributions are:
\begin{enumerate}[leftmargin=1.6em,itemsep=3pt,topsep=3pt,parsep=0pt]
\item Separate conditions for time advancement, state accumulation and output representation are derived and evaluated in diffusion, uniform-source and Robertson kinetics calculations. All eleven configurations selected before integration using sufficient bounds satisfy their accuracy targets.
\item Analytical representation requirements for pressure differences, pressure reconstruction from conserved variables and linear solves identify how the magnitude of a background quantity, cancellation and the direction of a perturbation affect the required precision.
\item The \emph{Precision Atlas for Scientific Kernels} records the domains, representations, arithmetic roles and tolerances of these bounds alongside finite precision evaluations of 60 representative kernels. Analytical guarantees are distinguished from observed passing widths.
\item A probability model connects the rounding of a sensor to the joint selection of refinement blocks and cell-update work. Scalar transport, shock--vortex interaction and Rayleigh--Taylor flow test its consequences for evolving meshes and physical quantities.
\item A refinement criterion informed by the Atlas uses local sensor and pressure reconstruction bounds to exclude extra tags. Direct and Atlas-guided tagging are compared across sensor precisions and mesh resolutions, including combined state and sensor conversion. Loss of reference tags quantifies where the representation is inadequate.
\end{enumerate}

Section~\ref{sec:representation} formulates the representation and refinement requirements, describes the Atlas and specifies the numerical experiments. Section~\ref{sec:validation} reports the results for RQ1--RQ3: precision requirements, the effects of reduced precision on refinement, and the application of the Atlas to AMR. Section~\ref{sec:discussion} discusses precision selection and its limitations. Section~\ref{sec:conclusion} summarizes the findings. The appendices and supplementary data contain supporting derivations and controls.

\Needspace{8\baselineskip}
\section{Formulation and methods}\label{sec:representation}

\subsection{Representation and accuracy requirements}

Let $X$ be a specified domain of physical inputs, $J:X\to\mathbb R^m$ a target quantity, $Q_p$ an input encoding with $p$ binary significand bits (including the leading bit), and $O_p$ the permitted output set. The target can be the mathematical discrete solution, a physical solution, a vector of observations, or a discrete label. The accuracy requirement is an acceptance set $\mathcal A_p(x)$; for a maximum-norm error tolerance $\varepsilon$ it is $O_p\cap\{v:\|v-J(x)\|_\infty\le\varepsilon\}$. For a required discrete decision it contains only the prescribed label. A representation capable of expressing the target is a different requirement from an implementation capable of computing it.

For a fixed finite ordered family $\mathcal P$ (here generally the integer significand widths 4 through 53 with eight exponent bits), define
\begin{align}
\mathcal F_{\rm rep}&=\{p\in\mathcal P:\exists A_p\in\mathcal C_p,\ \forall x\in X,\ A_p(Q_p(x))\in\mathcal A_p(x)\},\\
\mathcal F_{\rm impl}&=\{p\in\mathcal P:\forall x\in X,\ \widehat A_p(Q_p(x))\in\mathcal A_p(x)\},\label{eq:feasible}
\end{align}
where $\mathcal C_p$ is the class of deterministic algorithms considered, and $\widehat A_p\in\mathcal C_p$ is the specified implementation. Then $\mathcal F_{\rm impl}\subseteq\mathcal F_{\rm rep}$. For either set $\mathcal F$ we distinguish
\begin{align}
p_{\rm first}(\mathcal F)&=\min\mathcal F,\\
p_{\rm sustained}(\mathcal F)&=\min\{p\in\mathcal P: \{q\in\mathcal P:q\ge p\}\subseteq\mathcal F\}.\label{eq:sustained}
\end{align}
A sustained threshold requires all higher admitted widths to pass. Finite tests of particular inputs define observed passing sets; a uniform requirement concerns the complete domain $X$. In named-format comparisons the exponent range must also be specified, since the ranges of BF16 and FP16 differ.

For the first feasible width, the necessary--sufficient relation has the following meaning. If every width smaller than $p_{\rm N}$ has been excluded for the representation class and the specified implementation is certified at $p_{\rm S}$, then
\begin{equation}
p_{\rm N}\le p_{\rm first}(\mathcal F_{\rm rep})\le p_{\rm first}(\mathcal F_{\rm impl})\le p_{\rm S}.
\label{eq:interval}
\end{equation}
The same bound on the sustained width requires certification of the entire upper tail. 
Appendix~\ref{app:decisions} gives a threshold predicate with an isolated passing width followed by a failing interval.

The representation exclusion follows from information-based complexity~\cite{Traub1988}. Two inputs with identical encodings and known data (such as coefficients) must have the same deterministic output. If $Q_p(x_1)=Q_p(x_2)$ but $\mathcal A_p(x_1)\cap\mathcal A_p(x_2)=\varnothing$, no such output is possible. An unrestricted deterministic decoder exists if the acceptance sets intersect over every encoding fiber. For bounded vector targets and unrestricted real output in maximum norm, the minimal worst-case error on a fiber is half its diameter, attained by the coordinatewise midpoints. For restricted outputs the intersection must also contain an allowed word.

For a fixed implementation of a time-dependent numerical method, write $x_{n+1}=\Phi_n(x_n)$ for the exact discrete map and $D_n$ for the decoded numerical state. The local defect $\eta_n=\|D_{n+1}-\Phi_n(D_n)\|$ and a Lipschitz bound $L_n$ for the exact map on the relevant domain give
\begin{equation}
E_{n+1}\le L_n E_n+\eta_n,\qquad E_n=\|D_n-x_n\|.
\label{eq:propagation}
\end{equation}
Iteration bounds the propagated implementation error. The full error budget includes discretization, initialization, temporal alignment and output evaluation. For adaptive trajectories that use different meshes, the maps $\Phi_n$ may differ; the error budget then depends on the changed discretization.

\subsection{Time advancement, state accumulation and output representation}\label{sec:time}

For a finite stored simulation time $z$ with finite successor, let $g_t(z)=\operatorname{succ}(z)-z$. If $H$ is the upper bound on the permitted realized duration, including any endpoint constraint, an advancing stored time exists exactly when $g_t(z)\le H$. For positive normal $z\in[2^{e_t},2^{e_t+1})$, $g_t=2^{e_t+1-p_t}$ and consequently
\begin{equation}
p_t\ge\left\lceil e_t+1-\log_2 H\right\rceil.
\label{eq:time-bits}
\end{equation}
The origin must also be representable and in range. A nearest-rounded request can stall or exceed the permitted duration even when an admissible step exists. Directed selection chooses the largest stored time not exceeding $z+H$ (Appendix~\ref{app:time}); storing an origin and elapsed time separately is another option when the required physical time arguments remain accessible.

For periodic diffusion on a unit interval with $N$ points and diffusivity $\kappa=1$, the centered semidiscrete operator has the forward-Euler ceiling $H_N=1/(2N^2)$. In the supported positive normal range, the first forbidden starting-time binade begins at $2^{p_t+\lfloor\log_2 H_N\rfloor}$.

Doubling resolution at a fixed normal time increases the spacing requirement by two bits. More generally, if a refinement level $\ell$ has spacing $h_\ell=h_0/2^\ell$ and a stability limit $H_\ell\propto h_\ell^a$, Eq.~\eqref{eq:time-bits} increases the time-width requirement by $a$ bits per level at fixed $e_t$ when $a$ is an integer. The corresponding exponent is one for a hyperbolic CFL limit at fixed wave speed and two for explicit diffusion. Smaller steps also reduce state increments. The time integration controls in Section~\ref{sec:time-results} isolate these representation effects. The spatial sensor condition is derived separately in Section~\ref{sec:tag-probabilities}.

The same timestep restriction affects state accumulation. For a stored value $s$ and incoming exact increment $d$, RN absorbs the change when $-(s-\operatorname{pred}(s))/2<d<(\operatorname{succ}(s)-s)/2$, with midpoint ties determined by parity. A separate constraint applies to the reported output: for a scalar or componentwise maximum-norm target $J_i$ reported in a specified format $F_o$, $\max_i\operatorname{dist}(J_i,F_o)$ is the smallest possible output error, regardless of internal arithmetic.

\subsection{Construction of the Precision Atlas}\label{sec:atlas}

The Precision Atlas contains the precision evidence for 60 representative numerical kernels. Each is evaluated on two inputs using 54 custom or named format choices, giving 6,480 evaluations. The 60 programs and their reference types are listed in Appendix~\ref{app:inventory}; 34 of the 120 scenarios use the same discrete algorithm in FP64 as a reference for arithmetic error. An Atlas entry specifies the domain, known data, output set, arithmetic roles, rounding policy and tolerance, and distinguishes analytical exclusions, sufficient conditions and finite passing sets.

The Lean 4 proofs~\cite{deMouraUllrich2021Lean4} establish rational successor feasibility and selection, nearest-cell geometry, normal-binade output bounds, and nonexpansivity and residual propagation for mathematical Euler updates. The formal evidence also includes a configured FP32 decoding and adjacency connection to FloatLib~\cite{GeorgeAdkissonAnandkumar2026FloatLib}, a verified floating-point arithmetic library. The FloatLib arithmetic environment uses Lean v4.35.0. 
The study-specific proof sources and verification records are part of the Precision Atlas. These results concern the stated arithmetic and lattice assumptions; they do not verify the full numerical implementation, the AMR decision rule or general nonlinear PDE trajectories.

For the fluid calculations of interest, pressure differences and pressure reconstruction from conserved variables determine the accuracy of the refinement sensor. Two small linear systems with the same condition number illustrate a complementary dependence on the direction of a perturbation, as described by effective conditioning~\cite{ChanFoulser1988}. These examples specify the information available at the input to each operation and the corresponding output accuracy.

Table~\ref{tab:matched} gives exact widths for seven controlled representation problems with absolute tolerance $2^{-q}$ and the common eight-exponent-bit family. The input of a fluctuation or internal-energy representation is available before quantization. For the linear examples, the matrix is known and exact; for pressure reconstruction the finite arithmetic and candidate output are specified. Appendix~\ref{app:matched} derives the collision witnesses (pairs of inputs that round to the same stored value but require different outputs, which excludes that width) and sufficient bounds, including the restricted-output centered-gradient width $q+2$. Supporting decision controls appear in Appendix~\ref{app:decisions}.

\begin{table}[H]
\centering\small
\caption{Exact first and sustained widths within the candidate family for the seven analytical examples. The two widths agree for these particular problems. $e_P$ denotes the absolute-pressure binade; $k$ is the kinetic-energy or matrix scale exponent; $q$ sets the tolerance. The last column illustrates the most demanding specified scale at the tighter tolerance. The fixed-LU implementation is a separate calculation.}
\label{tab:matched}
\begin{tabularx}{\linewidth}{@{}l X c c@{}}
\toprule
Case & Representation and target & Width & Example \\
\midrule
G1 & Absolute pressure; gradient & $e_P+5+q$ & 51 \\
G2 & Pressure fluctuation; gradient & $q+1$ & 15 \\
E1 & Total energy; gas pressure & $k+q-1$ & 47 \\
E2 & Internal energy; gas pressure & $q$ & 12 \\
L1 & Diagonal system; difference direction & $q$ & 12 \\
L2 & Rotated system; difference direction & $k+q$ & 44 \\
L3 & Rotated system; common direction & $q$ & 12 \\
\bottomrule
\end{tabularx}
\end{table}

\Needspace{6\baselineskip}
\subsection{Mathematical formulation of precision-dependent adaptive mesh refinement}\label{sec:theory}

\subsubsection{Threshold decisions under rounding}\label{sec:threshold-method}

Let $s_i$ be a scalar sensor on a mesh, and $\mathcal N(i)$ the set of distinct face-neighbor cells of $i$. Boundary copies are identified with the original cell, not independently rounded values. Define the reference indicator and strict decision by
\begin{equation}
I_i=\max_{j\in\mathcal N(i)}|s_i-s_j|,\qquad D_i=\mathbf 1\{I_i>\tau\},\quad\tau\ge0.
\label{eq:indicator}
\end{equation}
In the sensor-input analysis each unique sensor value is rounded once, and the differences and threshold are then evaluated exactly, or in FP64. If $|\widehat s_i-s_i|\le\eta_i$, the reverse triangle inequality and the maximum give
\begin{equation}
|\widehat I_i-I_i|\le M_i,\qquad M_i=\max_{j\in\mathcal N(i)}(\eta_i+\eta_j).
\label{eq:margin}
\end{equation}
A sufficient condition for a tag is $\widehat I_i-M_i>\tau$, and for a no-tag is $\widehat I_i+M_i\le\tau$. Ambiguous cases can be evaluated with a higher-precision filter~\cite{Shewchuk1997,Titolo2019}. On a common normal lattice of spacing $g$, the input-only bounds are $M_i\le g$ for RN and $M_i\le2g$ for SR. Near binade or range transitions the one-sided distances to neighboring words are used. Any rounding of the indicator or threshold must also enter the margin.

RN preserves a uniform sensor. For a nonuniform edge with true absolute difference $\delta$, suppose $0\le\delta\le\tau<g$ on a common lattice away from its boundaries. Under a uniform distribution of the pair's phase modulo $g$, the probability of a rounded jump $g$ is $\delta/g$: the phases crossing an RN boundary have length $\delta$, and all other phases give zero difference. This probability describes random input alignment. The four edge events of a stencil share the center value; for a fixed field the decisions can be evaluated directly.

\subsubsection{Cell-tagging probabilities}\label{sec:tag-probabilities}

For an exact finite value $s_i$ bracketed by adjacent representable words $a_i,b_i$, ideal SR gives $b_i$ with probability $\theta_i=(s_i-a_i)/(b_i-a_i)$ and $a_i$ otherwise. Exactly represented values are unchanged. Its zero conditional mean error does not imply that a nonlinear indicator is unbiased~\cite{Connolly2021,CrociEtAl2022}. Let $S_i$ be the finite support and $\pi_i(z)$ the mass of the rounded value $z$. Under independent rounding of distinct cell values, conditioning on the central word gives the exact probability
\begin{equation}
P_i=1-\sum_{z\in S_i}\pi_i(z)
\prod_{j\in\mathcal N(i)}\left[\sum_{w\in S_j}\pi_j(w)\,\mathbf 1\{|w-z|\le\tau\}\right].
\label{eq:tag-general}
\end{equation}
For a frozen field, expected false-positive and false-negative counts follow by linearity of expectation:
\begin{equation}
\mathbb E N_{\rm FP}=\sum_{i:D_i=0}P_i,\qquad
\mathbb E N_{\rm FN}=\sum_{i:D_i=1}(1-P_i).
\label{eq:fpfn}
\end{equation}
Expected counts are sums of marginal probabilities, irrespective of dependence among the tag events. The independence assumption in Eq.~\eqref{eq:tag-general} is on the distinct cell draws, with each rounded cell reused throughout the stencil. The equation applies across rounding intervals.

For a uniform field $s_i=c=a+\theta g$ with $g=b-a$, $0\le\theta\le1$, the signed rounded difference has zero expectation, whereas $\mathbb E|\widehat s_i-\widehat s_j|=2g\theta(1-\theta)$. On a uniform lattice the two-point distribution can equivalently be obtained by adding uniform dither on $[-g/2,g/2]$ before RN~\cite{GrayStockham1993}. If $g>\tau$, an interior cell with four distinct face neighbors remains untagged precisely when all five rounded words coincide. The probability of a false tag is therefore
\begin{equation}
P_{\rm uniform}=\begin{cases}
1-\theta^5-(1-\theta)^5,&g>\tau,\\
0,&g\le\tau.
\end{cases}
\label{eq:uniform}
\end{equation}
Equation~\eqref{eq:uniform} is the probability that a phantom gradient tags the cell.
At an exactly representable value the probability is zero; at a midpoint with $g>\tau$ it is $15/16$. At $\tau=g$ the strict comparison gives no tag. RN and common-variate SR preserve a uniform field; the latter introduces dependence between the cell draws, so its nonuniform-field probability requires the joint distribution.

For all uniform values in a normal binade of exponent $e_s$ with finite bracketing words, absence of this particular mechanism is equivalent to
\begin{equation}
g_p=2^{e_s+1-p}\le\tau,
\qquad p\ge\left\lceil e_s+1-\log_2\tau\right\rceil\quad(\tau>0).
\label{eq:spacing}
\end{equation}
Sufficiency follows because no encoded jump exceeds $g_p$. Necessity follows by choosing a midpoint with $g_p>\tau$; it also follows for any uniform local risk limit smaller than $15/16$. When $\tau=\Gamma_s h$, with fixed gradient threshold $\Gamma_s$ and cell spacing $h$, halving $h$ requires one additional significand bit at fixed $e_s$ to eliminate the uniform-field mechanism. Relative or normalized indicators require their own threshold and error model. Nonuniform decisions are governed by Eq.~\eqref{eq:margin}.

\subsubsection{Refinement-block probabilities}\label{sec:blocks}

The two-level policy tags base cells by Eq.~\eqref{eq:indicator}, adds one face-neighbor buffer, and refines each aligned square parent block intersecting the buffered tags. For a block $B$ of $b\times b$ parent cells, let $A_B=B\cup\bigcup_{i\in B}\mathcal N(i)$ be the set of potential original tags that would refine it. Let $V_B$ contain $A_B$ and all their distinct sensor neighbors, and let $E_B$ be the undirected edges joining each $i\in A_B$ to its sensor neighbors. At the domain boundary, periodic or clamped neighbor identities define these sets. A block remains unrefined precisely when every comparison on $E_B$ is at or below the threshold. Its conditional probability of refinement is
\begin{equation}
P_B=1-\sum_{\mathbf z\in\prod_{i\in V_B}S_i}
\left(\prod_{i\in V_B}\pi_i(z_i)\right)
\left(\prod_{(i,j)\in E_B}\mathbf 1\{|z_i-z_j|\le\tau\}\right).
\label{eq:block-general}
\end{equation}
The sum enforces every edge constraint on the same assignment of words. For an interior block in two dimensions, two face-neighbor expansions give $|V_B|=b^2+8b+4$: the block has $b^2$ cells, its first halo adds $4b$, and the second expansion adds $4b+4$. In a uniform field with $g>\tau$, all edges require equality of the rounded words. Since the graph is connected, the no-refinement event requires all $|V_B|$ words to coincide. Hence
\begin{equation}
P_B=1-\theta^{b^2+8b+4}-(1-\theta)^{b^2+8b+4}.
\label{eq:block-uniform}
\end{equation}
The block probability is zero for $g\le\tau$. For the $4\times4$ parent blocks used here, the exponent is 52. A relatively small tag probability can thus produce a substantial probability of selecting a block. Boundary blocks and general nonuniform fields are evaluated using Eq.~\eqref{eq:block-general}.

The joint sum is evaluated by transferring partial probabilities across rows of the neighbor graph. For the interior and clamped-boundary blocks studied here, the graph has width at most eight, with at most $2^8$ bracket selections per row. Horizontal constraints determine each row weight; vertical constraints give a compatibility matrix between rows. Transferring these weights row by row evaluates the finite sum without enumerating all $2^{52}$ assignments. The computation uses FP64 and is checked against exhaustive small-field enumerations.

\subsubsection{Mesh resolution and numerical work}\label{sec:work}

Suppose a two-level mesh in $d$ dimensions has $N_0$ equal-volume base cells and refinement ratio $r$. Its $N_B$ aligned blocks have equal size and tile the base mesh. Let $A$ be the fraction of base cells refined (equivalently the fraction of blocks refined). The realized number of active leaves and its expectation conditional on the sensor field are
\begin{align}
N_{\rm leaf}&=N_0[1+(r^d-1)A],\\
\mathbb E[A\mid\mathbf s]&=\frac1{N_B}\sum_B P_B,\\
\mathbb E[N_{\rm leaf}\mid\mathbf s]&=N_0\left[1+\frac{r^d-1}{N_B}\sum_B P_B\right].\label{eq:expected-leaves}
\end{align}
No independence of block decisions is needed for this mean. If each active leaf receives $K_\ell$ updates on the mesh selected at regrid $\ell$, the work is
\begin{equation}
W=N_0\sum_\ell K_\ell[1+(r^d-1)A_\ell].
\label{eq:work}
\end{equation}
For a predetermined number of updates between regrids, the conditional expected work follows from Eq.~\eqref{eq:expected-leaves}. Over a trajectory, the distribution of later fields depends on the earlier rounding and refinement events; expected total work requires averaging over that evolving distribution.

Together Eqs.~\eqref{eq:tag-general}, \eqref{eq:block-general}, \eqref{eq:expected-leaves} and \eqref{eq:work} give a conditional mathematical connection from floating-point representation to threshold decisions, realized spatial resolution and numerical work. The results in Section~\ref{sec:uniform-validation} and the subsequent flow comparisons examine these links before considering the physical prediction.

\subsubsection{Pressure reconstruction from conserved variables}\label{sec:pressure-method}

The total-energy example in Table~\ref{tab:matched} specifies a precision requirement upstream of the pressure sensor. For conserved variables $U=(\rho,\boldsymbol m,E)$, with momentum density $\boldsymbol m$ and total-energy density $E$, ideal-gas pressure is $P(U)=(\gamma-1)(E-\|\boldsymbol m\|^2/(2\rho))$. For $\rho,\rho+\delta\rho>0$, the exact change due to a state perturbation is
\begin{equation}
\begin{split}
P(U+\delta U)-P(U)=(\gamma-1)\bigg[\delta E
-\frac{2\boldsymbol m\cdot\delta\boldsymbol m+\|\delta\boldsymbol m\|^2}{2(\rho+\delta\rho)}
+\frac{\|\boldsymbol m\|^2\delta\rho}{2\rho(\rho+\delta\rho)}\bigg].
\end{split}
\label{eq:pressure-state}
\end{equation}
The pressure perturbation follows from the state error, to which pressure-evaluation and any additional sensor-conversion errors contribute. For a uniform state with independently converted cells and common sensor support $S$ of positive probability masses, the necessary and sufficient condition for zero false-tag probability is $\operatorname{diam}(S)\le\tau$. If the diameter exceeds $\tau$, adjacent cells can attain the endpoints with positive probability. Equation~\eqref{eq:spacing} is the direct two-word special case. On nonuniform fields an error envelope can be propagated into Eq.~\eqref{eq:margin}, or the composed distribution used in Eq.~\eqref{eq:tag-general} when its independence assumptions hold. Correlations introduced by remapping and subsequent evolution require a joint model.

\subsubsection{Atlas-guided refinement criterion}\label{sec:guided-method}

For a nonuniform field let $\widehat s_i$ be the stored sensor and $\eta_i$ an upper bound on $|s_i-\widehat s_i|$. Direct tagging uses $\widehat I_i>\tau$. Using Eq.~\eqref{eq:margin}, the Atlas-guided criterion instead uses
\begin{equation}
\widehat I_i>\tau+M_i,\qquad M_i=\max_{j\in\mathcal N(i)}(\eta_i+\eta_j).
\label{eq:atlas-guided-tag}
\end{equation}
If $\widehat I_i+M_i\le\tau$, the cell is below threshold. Otherwise, unless Eq.~\eqref{eq:atlas-guided-tag} holds, the decision is unresolved; we leave such cells untagged. The bounds use the neighboring representable words (half-cell distances for RN and full inverse intervals for SR), with outward rounding. Buffering and block selection remain unchanged. The rule excludes extra tags, not all changes to the mesh: it can also miss reference tags. The Atlas itself is unchanged and need not be queried during the calculation.

\textbf{State and sensor conversion.} The bound on pressure must include errors in its inputs. Let $(r,\boldsymbol m,E)$ be the restricted decoded conserved state, and let $\eta_r,\eta_{m,k},\eta_E$ enclose the corresponding exact means before conversion. For a positive lower density bound $r_{\min}$, Eq.~\eqref{eq:pressure-state} gives the pressure radius
\begin{equation}
\eta_P=(\gamma-1)\left[\eta_E+
\frac{2\sum_k |m_k|\eta_{m,k}+\sum_k\eta_{m,k}^2}{2r_{\min}}+
\frac{\|\boldsymbol m\|^2\eta_r}{2rr_{\min}}\right].
\label{eq:atlas-pressure-radius}
\end{equation}
The E1 pressure reconstruction analysis supplies this dependence; E2 illustrates the alternative internal-energy representation. The G1/G2 spacing analysis supplies the sensor-conversion radius. We use $\eta_i=\eta_P+\eta_{\rm sensor}+\eta_{\rm arith}$, with the latter term bounding restriction, primitive-variable reconstruction and comparison arithmetic. Component intervals are propagated from the active leaves through conservative restriction; the implementation and a comparison of the two radii are given in the technical supplementary data. If the lower density bound is not positive or the radius is unbounded, the cell is not tagged.

\subsubsection{Atlas entry for refinement tagging}\label{sec:atlas-tag-entry}

Table~\ref{tab:atlas-entry} gives the entry for the AMR sensor considered here. Its requirement concerns the elimination of uniform-field false tags. The block-selection probabilities relate the local decision to the selected resolution. Section~\ref{subsec:atlas-amr-application} uses the G1/G2 and E1/E2 analytical bounds from Table~\ref{tab:matched} in the tag criterion; the finite passing widths do not enter that rule.

\begin{table}[!htb]
\centering\small
\caption{Example Precision Atlas entry for the strict face-neighbor tagger. The entry concerns elimination of uniform-field false tags. The electronic Atlas entry records the representation, decision rule, mesh policy, assumptions and supporting evidence.}
\label{tab:atlas-entry}
\begin{tabularx}{\linewidth}{@{}p{.23\linewidth}X@{}}
\toprule
Field & Value / scope \\
\midrule
Target and domain & No refinement tag on any uniform field in a positive normal binade $[2^{e_s},2^{e_s+1})$ with supported finite brackets. \\
Arithmetic role & Sensor encoding; exact neighbor-difference comparison; no state rounding. \\
Policy and family & Independent ideal distance-proportional SR, fixed exponent range, significand width $p$. \\
Decision and tolerance & Strict $I_i>\tau$; require probability of a false tag equal to zero for every admissible uniform value; $\tau>0$. \\
Feasible set & All in-range widths with $2^{e_s+1-p}\le\tau$ (first and sustained agree on this family). \\
Mesh rule & One face-cell buffer and aligned $4\times4$ parent blocks, two levels, refinement ratio two. \\
Derived quantities & Conditional tag and block probabilities, expected refined area and leaf count, Eqs.~\eqref{eq:tag-general}--\eqref{eq:work}. \\
Evidence and limits & Uniform-field derivation; frozen shock--vortex ensemble; numerical trajectory comparisons. \\
\bottomrule
\end{tabularx}
\end{table}

\FloatBarrier
\subsection{Numerical experiments and reference solutions}\label{sec:experiments}

\subsubsection{Time integration experiments}\label{sec:time-experiments}

Consider a spatially uniform source $P(\xi)=1+\sigma\xi$, $0\le\xi\le L$, with $\sigma=2^{-8}$ and $L=2^{-9}$. The diffusion term vanishes, but the timestep is capped by the periodic diffusion ceiling with $\kappa=1$; each request is $3H_N/5$ with endpoint clipping. The use of a constant source isolates the storage mechanism analyzed by Croci and Giles~\cite{CrociGiles2023} and Barad~\cite{Barad2026} from the formation of the increments.

The nonlinear comparison uses the Robertson reaction system~\cite{Robertson1966}, with $a_R=1/25$, $b_R=10^4$, $c_R=3\times10^7$,
\begin{equation}
\dot y_1=-a_Ry_1+b_Ry_2y_3,\quad
\dot y_2=a_Ry_1-b_Ry_2y_3-c_Ry_2^2,\quad
\dot y_3=c_Ry_2^2.
\label{eq:robertson-ode}
\end{equation}
Two short restart windows, beginning at the FP32 predecessor of 4096 and at 4096, use common FP64 times, identical FP32 initial words and the same FP32 derivative/product operations. The derivative is evaluated on the projected state of each accumulation policy, allowing nonlinear feedback to differ. The frozen-Jacobian modal ceiling for explicit Euler is evaluated on numerical reference states obtained using Radau and backward differentiation formula (BDF) methods.

The selection procedure tests a fixed list of arithmetic configurations against the input domain and accuracy requirements before state integration. For periodic diffusion, candidate time representations are considered in the order absolute FP16, elapsed FP16, absolute FP32, elapsed FP32 and elapsed FP64, with FP64 state and output arithmetic. A candidate must admit an advancing schedule that reaches the observation times exactly; its bounds on implementation and physical error must satisfy the respective tolerances. For the uniform-source problem, the elapsed-time schedule is fixed in FP64, with FP32 incoming products. The accumulation candidates are plain FP32, compensated FP32 and FP64 accumulation, with permitted outputs tested in FP32-then-FP64 order. The first candidate satisfying both decoded-state and output error bounds is selected. If none qualifies, the procedure supplies no certified configuration. The analytical bounds and admissibility conditions are given in Appendix~\ref{app:time} and the technical supplementary data.

\subsubsection{Adaptive mesh refinement experiments}\label{sec:amr-experiments}

The reduced formats are bfloat16 (BF16), IEEE half precision (FP16) and IEEE single precision (FP32), with significand and exponent widths $(8,8)$, $(11,5)$ and $(24,8)$, respectively. IEEE double precision (FP64) has $(53,11)$; the custom significand family has eight exponent bits. The software implements gradual underflow and nearest-even ties. SR uses a 53-bit uniform random grid; the draws are indexed by seed, cell, regrid and field, so each realization is reproducible. Its upper-word probability is $\lceil2^{53}\theta_i\rceil/2^{53}$, within one random-grid step of the ideal value. A coupling of $m$ independent draws bounds the corresponding event-probability difference by $m2^{-53}$. In the mathematical model the distinct draws are independent. In the sensor-only comparisons, each unique sensor value is encoded once per regrid; state evolution, neighbor differences and threshold comparisons use FP64. The state-storage controls also use FP64 flux and update arithmetic. Both refinement levels use a common timestep without subcycling.

The scalar transport problem uses a conservative first-order upwind discretization of $q_t+q_x+q_y/2=0$ on the periodic unit square. The initial profile is $q(x,y,0)=\cos^8(\pi(x-1/4))\cos^8(\pi(y-3/8))$. Base grids have $N=16,32,64$ cells per direction, with a factor-two refinement level, timestep $1/(4N)$ and eight regrids before the final time $1/2$. The low-background sensor is $1.5+2^{-8}q$ with jump threshold $2^{-8}(0.75/N)$; direct fluctuation encoding uses $q$ and the corresponding rescaled threshold. Each primary SR case has 32 seeds. The volume-weighted $L_1$ error compares the leaf field with the analytical translated cell averages on the common finest lattice.

The compressible calculations solve the ideal-gas Euler equations with $\gamma=1.4$ and a passive conservative marker. A two-level finite-volume scheme uses piecewise-constant states, Harten--Lax--van Leer (HLL) fluxes and forward Euler. Each face segment carries one flux, multiplied by its length and applied with opposite signs to the neighboring volume-integrated states; at a coarse--fine interface, the two fine-face contributions are summed in the coarse-cell update. In the Rayleigh--Taylor (RT) problem, the density plateaus are 1 and 2, the Atwood number is $1/3$, the gravity is unity and the initial interface thickness is $1/64$ (detailed in Appendix~\ref{app:cfd}). Density is used as the native sensor. The base meshes are $32\times128$ and $64\times256$ through $t=1$. To follow the nonlinear instability, a longer calculation on these meshes and on $128\times512$ is performed through $t=3.75$, with regrids every $1/8$ and a gradient threshold $\tau=2h$, $h=1/N_y$. The shock--vortex (SV) configuration is motivated by the Jiang--Shu test~\cite{JiangShu1996}: a Mach-1.1 normal shock at $x=1/2$ in $[0,2]\times[0,1]$, upstream density and pressure one, and a vortex of strength 0.3 centered at $(1/4,1/2)$ with core radius 0.05. The downstream state follows the normal-shock relations. Pressure is used as the native sensor, with $\tau=0.09h$. The two base meshes are $128\times64$ and $256\times128$ with eight regrids through $t=0.6$. All CFD calculations use aligned $4\times4$ parent blocks and the same timestep schedule for all rounding policies at a given base resolution; the CFL cap is 0.4. Appendix~\ref{app:cfd} specifies the RT flux correction, conservative remapping, and initial and boundary conditions.

The primary native-sensor SV and short-time RT SR comparisons use 16 seeds per format and resolution. The late RT spatial comparison in Section~\ref{sec:rt-results} uses one BF16 SR realization; the tagging comparison in Section~\ref{subsec:atlas-amr-application} includes four SR seeds (0--3) at each grid, with the same FP64 and BF16 RN references. RN is deterministic. For sensor-only tests, the reference is the unrounded FP64 sensor on the same state. The extra-tag fraction counts candidate-only decisions divided by all base-cell decisions across regrids. Realized area is the fraction of base cells refined after buffering and block selection. Work is the cumulative number of active leaf-cell updates over all timesteps. For both RT and SV, uniform FP64 calculations with mesh spacings $h$ (uniform coarse), $h/2$ (same-finest uniform) and $h/4$ (twice-finest uniform), together with FP64 adaptive calculations with two refinement levels of spacings $h$ and $h/2$, provide discretization comparisons through $t=1$ and $t=0.6$, respectively.
Reference spread (defined as the variation in a physical observable obtained from FP64 calculations at different spatial resolutions and timesteps) is used to assess whether differences in physical observables are resolved. Confidence intervals are formed from run or frozen-field means, since cells share stencil inputs.

The coarse SV calculation with FP16 conserved-state conversion at initialization and after each timestep, followed by FP16 pressure-sensor conversion, tests the composed bound. The reference here is the pressure from the current leaves immediately before their most recent state conversion, not the FP64 pressure of already rounded leaves or an independently evolved FP64 trajectory.

\FloatBarrier
\section{Results}\label{sec:validation}

\subsection{RQ1: Precision requirements of scientific operations}

\subsubsection{Time advancement, state accumulation and output representation}\label{sec:time-results}

The periodic diffusion and uniform-source controls test the spacing, accumulation and output requirements in Section~\ref{sec:time} using the procedures in Section~\ref{sec:time-experiments}. For periodic diffusion, the timestep ceiling is $H_N=1/(2N^2)$. For FP16 the first forbidden starting-time binade moves from 1 at $N=32$ to $1/64$ at $N=256$. Forcing a successor at twice the ceiling gives the alternating mode a multiplier $-3$, even though a smooth trajectory can temporarily meet its error target.

\begin{figure}[!htb]
\centering
\includegraphics[width=.84\linewidth]{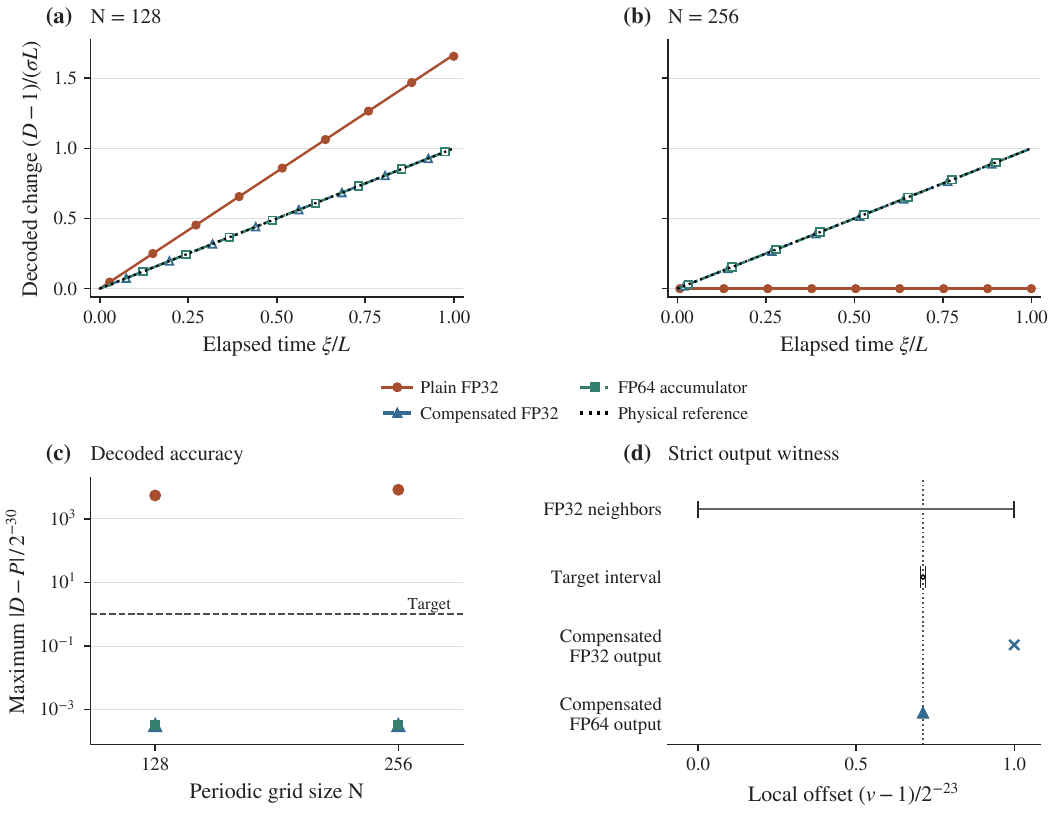}
\caption{\textbf{Source accumulation and output resolution.} (a,b) Decoded changes for $P(\xi)=1+2^{-8}\xi$, normalized by the exact final change, on the two periodic diffusion timestep schedules. (c) Decoded error for $N=128,256$, $\kappa=1$. (d) Output-format comparison at $N=96$, $\kappa=3/2$, tolerance $2^{-30}$. All accumulation policies receive the same FP32 products. Here $\sigma=2^{-8}$, $L=2^{-9}$ and $\xi$ is elapsed time.}
\label{fig:source}
\end{figure}

At $P=1$ the FP32 spacing is $2^{-23}$. The ordinary source increment is 0.60 spacings at $N=128$ and 0.15 at $N=256$. Plain FP32 accumulation produces 106 movements in 107 steps at the lower resolution, giving a final change 42 spacings above the exact change, and absorbs all 427 increments at the higher resolution. Compensated FP32 and FP64 accumulation of the same FP32 products give maximum decoded errors of about $3.0\times10^{-13}$ (Fig.~\ref{fig:source}).

In the exact uniform-source example, the target interval lies within $[1,2)$, so $p=30$ bounds the output error by $2^{-30}$. Saved targets exclude $p=29$ at the same tolerance, and the smaller output lattices are subsets of the 29-bit lattice. Therefore, both histories require a fixed output width of 30 bits; the final value $1+2^{-17}$ is exactly representable in 18 bits and fails the tolerance at all smaller widths. Unlike the general algorithmic feasible sets in Eq.~\eqref{eq:feasible}, these output-only word sets are nested. The result gives a reason to specify the observation set of a precision requirement. It also identifies situations in which improved internal arithmetic cannot meet a requested output tolerance.

The prospective tests (selected before integration) comprise four diffusion and eight source cases. The eleven selected configurations satisfy all prescribed targets. In the remaining source case, only FP32 output is allowed at tolerance $2^{-30}$, and the output lattice excludes acceptable words at 88 prescribed observation endpoints despite an exactly representable final value. Five independently executed comparison configurations meet their targets even though their sufficient error bounds exceed the applicable tolerances. Matched third-order Runge--Kutta (RK3) and conjugate-gradient (CG) calculations are given in Appendix~\ref{app:rk-cg}.

\subsubsection{Robertson kinetics and output representation}\label{sec:robertson}

The Robertson tests in Section~\ref{sec:time-experiments} examine the corresponding time-spacing and state-accumulation requirements in a nonlinear reaction system.

In the sampled FP32 time search, the spacing crosses the ceiling between adjacent times near $3.87\times10^3$, whereas nearest addition of the requested cap stagnates at 4096. A separate scalar late-mode calculation at $T=10^{11}$ gives time-spacing requirements of 49 bits for classical fourth-order Runge--Kutta (RK4) and 50 for Euler (Appendix~\ref{app:time}).

\begin{figure}[!tb]
\centering
\includegraphics[width=.88\linewidth]{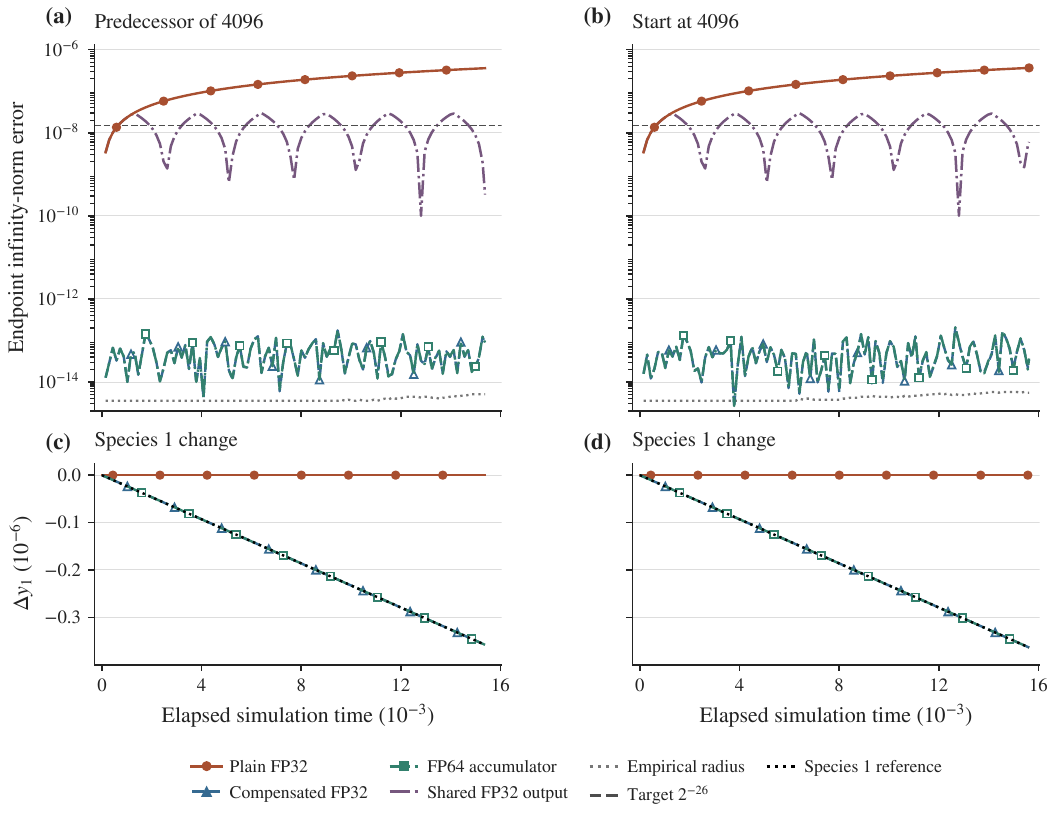}
\caption{\textbf{State accumulation and reporting in Robertson kinetics.} Panels (a,b) show infinity-norm errors at saved endpoints for restart windows at the FP32 predecessor of 4096 and at 4096; panels (c,d) show the signed species-1 changes. All arms use FP64 time and FP32 derivative/product operations. Compensated FP32 and FP64 accumulation retain small changes that plain FP32 loses. The FP32 reported output remains subject to the output lattice. All nonlinear physical comparisons are relative to numerical references with empirical resolution checks.}
\label{fig:robertson}
\end{figure}

Plain FP32 leaves species 1 unchanged, producing maximum full-state differences from the numerical reference of $3.58\times10^{-7}$ and $3.64\times10^{-7}$, equal to the species-1 changes over the two windows. Compensated FP32 and FP64 accumulation of FP32 products reduce these differences to $1.75$--$2.08\times10^{-13}$, with overlapping empirical reference-error intervals. Their FP32 reports exceed the intermediate-time target $2^{-26}$ (Fig.~\ref{fig:robertson}). For species 2, the compensated maximum error amounts to 7.85\% and 9.24\% of the maximum physical change, respectively. Thus the precision required to resolve changes in one species can differ from that required by the full-state absolute-error criterion.

\subsubsection{Precision requirements of controlled scientific quantities and the Atlas}\label{sec:atlas-results}

The seven analytical cases in Table~\ref{tab:matched} and the numerical kernel evaluations in Section~\ref{sec:atlas} examine how representation, physical scale, perturbation direction and tolerance affect the required precision.

A pressure background $B_P=3\,2^{e_P-1}$ leaves the exact gradient unchanged, but the required width grows with $e_P$ when absolute pressures are encoded. Similarly, with kinetic energy $2^k$, the pressure reconstruction error due to total-energy storage is proportional to $2^{k-p}$. The fluctuation and internal-energy representations remove these scale dependences in the specified domains. In the AMR calculation, the attainable pressures from rounded conserved words determine the sensor distribution (Section~\ref{sec:state-sensor}). For the linear systems with condition number $2^{32}$, the three right-hand-side representation minima are 12, 44 and 12 bits at tolerance $2^{-12}$. The corresponding first passing widths for pivoted LU on 32 fixed inputs are 12, 44 and 43. Across both inputs, FP32 passes 36 of the 60 numerical representatives listed in Table~\ref{tab:inventory} at scaled tolerance $10^{-3}$ and 22 of 60 at $10^{-6}$ (defined in Appendix~\ref{app:rk-cg}).

\FloatBarrier
\Needspace{6\baselineskip}
\subsection{RQ2: Effects of reduced precision on adaptive mesh refinement}

\subsubsection{Verification of cell and block probabilities}\label{sec:uniform-validation}

The uniform-field controls test the analytical cell and block probabilities and the spacing condition derived in Sections~\ref{sec:tag-probabilities} and~\ref{sec:blocks}.

\begin{figure}[!tb]
\centering
\includegraphics[width=0.88\linewidth]{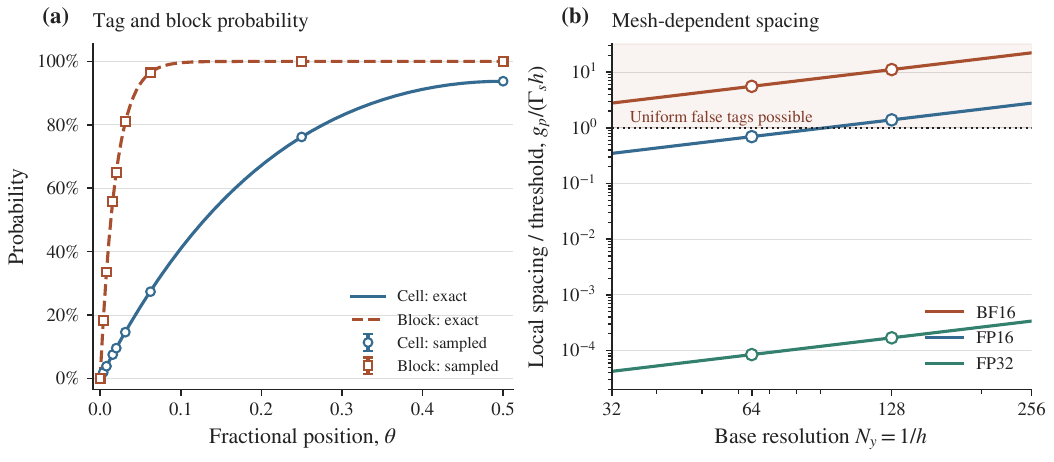}
\caption{\textbf{Local rounding and refinement resolution.} (a) Analytical cell and interior $4\times4$ block probabilities for a nonrepresentable uniform field at $\tau=g/2$. Points with 95\% field-level intervals show both cell tags and realized blocks from 512 independent periodic fields per fractional position. The symmetric half interval is displayed to resolve the intermediate block probabilities; the complete 16-position data are provided electronically.  (b) Ratio of spacing to the SV threshold $\tau=0.09h$ for the positive normal binade $e_s=0$. Open markers identify the two simulated base resolutions. The shaded region allows the uniform-field false-tag mechanism.}
\label{fig:probability}
\end{figure}

For BF16 midpoint rounding, 64 independent $32\times32$ fields at $\tau=g/2$ give a pre-buffer false-tag fraction of 93.57\% (95\% field-level interval 93.32--93.83\%), compared with the analytical value 93.75\%. RN and common-variate SR give zero false tags. To assess block selection away from saturation, we sample the same two-word distribution on periodic $32\times32$ meshes at 16 fractional positions in $[0,1]$, with 512 fields per position. The one-cell face buffer and aligned $4\times4$ blocks give 52 distinct inputs per block. At $\theta=0.02$, the cell and block probabilities are 0.096079 and 0.650251, respectively; the measured fractions are 0.095587 and 0.649567 (block interval [0.640948,0.658185]). Figure~\ref{fig:probability}a compares the full calculation with samples generated under its rounding distribution. This verifies the probability implementation and the amplification from the stencil to the block, including at intermediate probabilities.

In simulations initialized with a uniform state, encoding a nonrepresentable sensor with independent SR induces extra refinement while the physical state remains constant to within rounding.

\subsubsection{Shock--vortex interaction: fixed-field analysis}\label{sec:frozen}

The initial shock--vortex pressure field tests the conditional probability model of Sections~\ref{sec:tag-probabilities} and~\ref{sec:blocks} on a nonuniform sensor.

The normal-shock pressure provides a nonrepresentable sensor value without an artificial background. For $M_1=1.1$ and $\gamma=1.4$, $P_2=1+2\gamma(M_1^2-1)/(\gamma+1)=1.245$. Its adjacent BF16 words are 1.2421875 and 1.25, so $g=2^{-7}$ and $\theta=0.36$. Equation~\eqref{eq:uniform} gives $P_{\rm uniform}=0.8865792$ when $g>\tau$; the interior-block probability is close to unity. If the downstream pressure is approximated as uniform over $3/4$ of the initial domain, the contribution to the extra-tag fraction is 66.49\%. For FP16 on the finer grid, the corresponding cell probability is 47.22\%, or 35.42\% over that downstream fraction. Shock, vortex, boundary and diffusion effects motivate evaluation of the nonuniform pressure field.

We evaluate the nonuniform-field expressions on the initial FP64 pressure field at $128\times64$ using its decoded values, the strict threshold and the clamped neighbor identities. There are 8,192 cells and 512 possible blocks, with 520 reference tags and 64 reference blocks. RN masks are reconstructed directly; for SR the conditional expectations use the input brackets. The frozen-field comparison is a consistency check using independent draws from the model.

\begin{table}[!tb]
\centering\small
\caption{Initial SV pressure field ($128\times64$). Predictions are conditional expectations from Eqs.~\eqref{eq:fpfn} and \eqref{eq:block-general}; observations are means over 16 SR fields, with field-level 95\% intervals. The common unrounded sensor gives 520 tags and 64 realized blocks.}
\label{tab:frozen}
\begin{tabularx}{\linewidth}{@{}l r r X r r X@{}}
\toprule
&\multicolumn{3}{c}{Extra tags (count)}&\multicolumn{3}{c}{Refined blocks (count)}\\
\cmidrule(lr){2-4}\cmidrule(l){5-7}
Format & RN & SR pred. & SR mean [95\% CI] & RN & SR pred. & SR mean [95\% CI]\\
\midrule
BF16 & 16 & 5472.17 & 5455.25 [5433.69, 5476.81] & 70 & 447.33 & 446.75 [445.71, 447.79]\\
FP16 & 0 & 10.82 & 10.44 [9.01, 11.87] & 64 & 68.06 & 67.81 [67.08, 68.55]\\
FP32 & 0 & 0 & 0 [0, 0] & 64 & 64 & 64 [64, 64]\\
\bottomrule
\end{tabularx}
\end{table}

\begin{figure}[!tb]
\centering
\includegraphics[width=\linewidth]{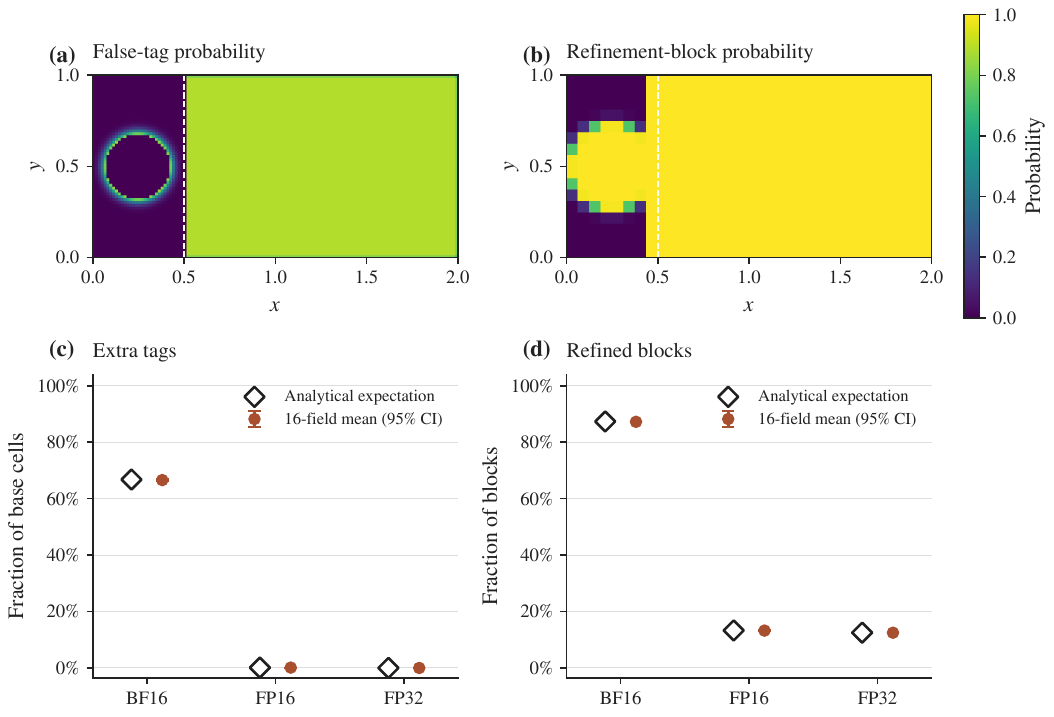}
\caption{\textbf{Conditional mesh-selection probabilities on a fixed nonuniform pressure field.} (a) BF16 ideal-SR false-tag probability, set to zero on cells already tagged by the unrounded reference. (b) Probability of refining each $4\times4$ parent block, accounting for all shared word values in its selection rule. Dashed lines locate the initial shock. (c,d) Expected extra tags and refined blocks, normalized by the 8,192 cells and 512 blocks, respectively, compared with the 16-field means and 95\% intervals. The analytical calculation uses the given pressure and the rounding model.}
\label{fig:frozen}
\end{figure}

The expected BF16 extra-tag count is $5.47\times10^3$, compared with the observed mean of $5.46\times10^3$ (Table~\ref{tab:frozen}). Both BF16 and FP16 expectations fall within the respective intervals. The predicted and observed mean BF16 block counts correspond to approximately $2.97\times10^4$ and $2.96\times10^4$ active leaves, respectively. The initial block probabilities are predominantly near zero and one (Fig.~\ref{fig:frozen}b). The downstream region dominates the BF16 extra tags in panel (a), and buffering extends the refined mesh beyond the individually tagged cells. The intermediate-probability tests in Fig.~\ref{fig:probability}a and Section~\ref{sec:regrid-check} exercise the probability evaluator away from saturation.

\subsubsection{Shock--vortex interaction: evolving mesh and numerical work}\label{sec:sv-results}

The evolving shock--vortex experiments in Section~\ref{sec:amr-experiments} test the consequences of the spacing condition and the work relation in Sections~\ref{sec:tag-probabilities} and~\ref{sec:work}.

In the downstream pressure binade the FP16 spacing is $2^{-10}=9.765625\times10^{-4}$. At $N_y=64$ and 128, the sensor thresholds are $1.40625\times10^{-3}$ and $7.03125\times10^{-4}$ (Fig.~\ref{fig:probability}b). The coarse threshold exceeds the spacing, whereas the finer threshold is smaller. The uniform-field condition consequently requires 11 and 12 significand bits on the two meshes, respectively. FP16 can generate uniform false tags on the finer grid.

\begin{figure}[!tb]
\centering
\includegraphics[width=\linewidth]{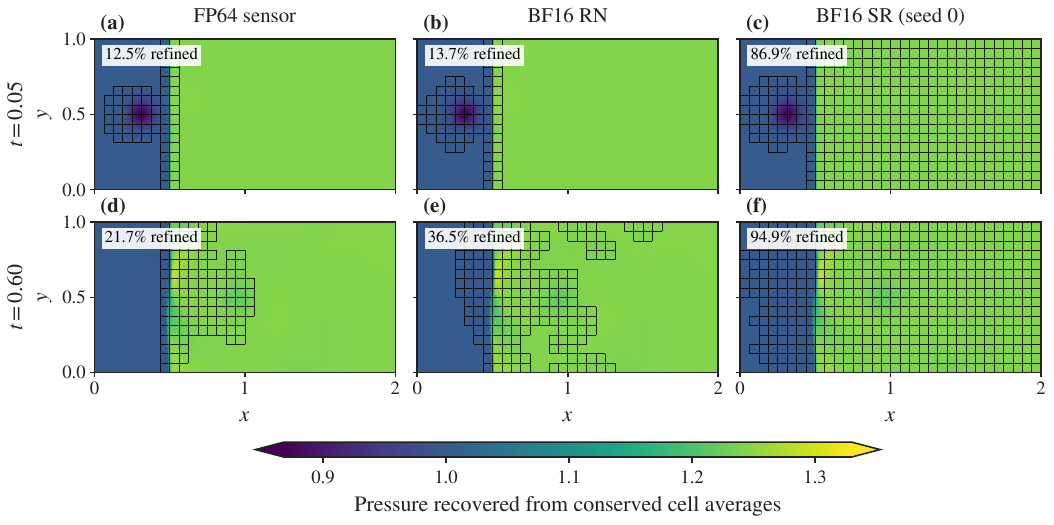}
\caption{\textbf{Spatial resolution selected by the rounded pressure sensor.} Shock--vortex interaction, base grid $128\times64$, factor-two refinement. The columns differ only in sensor encoding: FP64, BF16 RN and BF16 independent SR (seed 0). Colors show pressure and outlines mark refinement blocks at the indicated times.}
\label{fig:sv}
\end{figure}

Figure~\ref{fig:sv} shows the pressure and the resulting mesh at two times. At $t=0.05$, the FP64 reference, BF16 RN and BF16 SR refine 12.5\%, 13.7\% and 86.9\% of the domain, respectively; at $t=0.6$ the corresponding fractions are 21.7\%, 36.5\% and 94.9\%.

\begin{figure}[!tb]
\centering
\includegraphics[width=\linewidth]{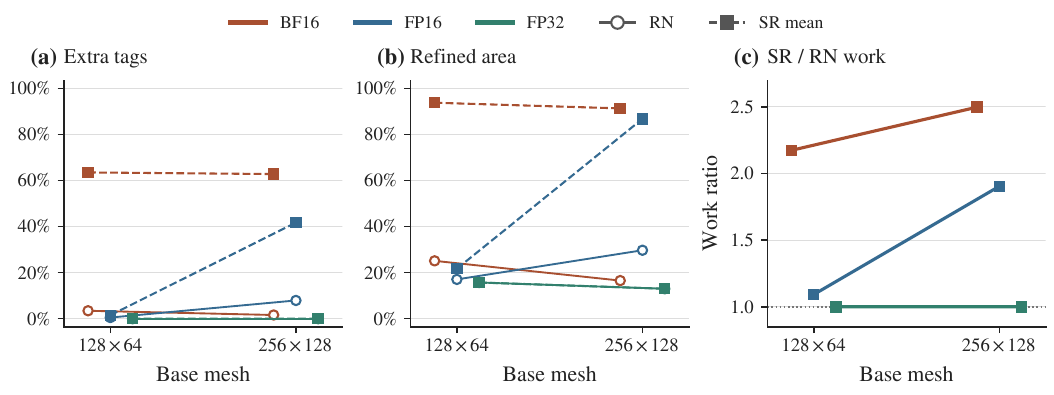}
\caption{\textbf{Precision-dependent mesh work under two spatial resolutions.} Extra-tag fractions, realized areas and SR/RN work ratios for the native pressure sensor. The SR uncertainty bars are the field/trajectory-level 95\% mean intervals across 16 seeds, with RN points. In panel (c) the ratios use the corresponding RN work as denominator. The work ratios follow from the realized areas and common schedule.}
\label{fig:work}
\end{figure}

Table~\ref{tab:work} and Fig.~\ref{fig:work} quantify the change from tags to realized refinement. On the finer mesh, FP16 RN refines 29.72\% compared with 13.01\% for FP32 RN, consistent with threshold crossings on nonuniform fields. The FP16 SR extra-tag fraction increases from 1.50\% on the coarse grid to 41.73\% on the finer grid. BF16 SR produces extensive additional tags on both meshes.

\begin{table}[!tb]
\centering\small
\caption{Simulated shock--vortex cases. Extra tags are candidate-only decisions divided by all base-cell decisions over regrids. Realized area is the update-weighted refined fraction, including buffering and blocks. SR entries are means over 16 seeds. Work ratios follow from the refined areas under the common step schedule (Eq.~\eqref{eq:work}).}
\label{tab:work}
\begin{tabular}{@{}llrrr@{}}
\toprule
Base mesh & Format & Extra tags RN / SR (\%) & Area RN / SR (\%) & $W_{\rm SR}/W_{\rm RN}$\\
\midrule
$128\times64$ & BF16 & 3.45 / 63.50 & 25.12 / 93.78 & 2.17\\
& FP16 & 0.50 / 1.50 & 17.07 / 21.71 & 1.09\\
& FP32 & 0 / 0 & 15.72 / 15.72 & 1.00\\
$256\times128$ & BF16 & 1.61 / 62.81 & 16.55 / 91.31 & 2.50\\
& FP16 & 7.93 / 41.73 & 29.72 / 86.78 & 1.90\\
& FP32 & 0 / $<0.001$ & 13.01 / 13.01 & 1.00\\
\bottomrule
\end{tabular}
\end{table}

With $d=2,r=2$, Eq.~\eqref{eq:work} reduces to $W=N_0K(1+3\overline A)$, where $K$ is the number of timesteps and $\overline A$ is the update-weighted refined area. The coarse BF16 pair with $K=960$ requires approximately $1.38\times10^7$ updates for RN and a mean of $3.00\times10^7$ for SR, yielding a ratio of 2.17. At the coarse RN area, the geometric maximum is $4/(1+3\overline A_{\rm RN})\simeq2.28$. The finer grid has a smaller RN area and therefore a different cap. The work increase follows from the additional selected cells.

\subsubsection{Conditional probabilities during mesh adaptation}\label{sec:regrid-check}

The conditional block probabilities of Section~\ref{sec:blocks} are evaluated on fields from successive regrids to test the model beyond the initial field. The fields come from six SV trajectories: the FP64 reference on both base grids, sensor-only SR for coarse BF16, coarse FP16 and fine FP16, and the coarse FP16 combined-state case. The fields immediately before remapping at $t=0,0.075,\ldots,0.525$ are used for the conditional calculation.

At each regrid Eq.~\eqref{eq:block-general} is evaluated on the fixed pressure field, and 64 fresh sensor realizations are generated on that field. Intervals are computed separately for each frozen field. Table~\ref{tab:conditional} gives pointwise 95\% intervals for the mean block count and for blocks with $0.05<P_B<0.95$. The maximum mean-area discrepancy is below 0.487 percentage points. For example, on the fine reference at $t=0.3$, the 233 intermediate-probability blocks have an expected mean of 0.491 and a sample mean of 0.497 (interval [0.486,0.507]). The tests condition on the state at each time and therefore assess the probability calculation separately from trajectory prediction. The expectations lie within 52 of 56 total-count intervals and 48 of 51 intermediate-block intervals; the four total-count misses are close to the 2.8 expected at nominal 95\% pointwise coverage. For the single intermediate block at $t=0.525$ in the combined-state sequence, the exact binomial interval contains the calculated probability.

\begin{table}[!tb]
\centering\small
\caption{Conditional checks at all eight regrids per sequence (64 sensor realizations per frozen field). Intermediate denotes $0.05<P_B<0.95$; the range gives the number of such blocks over the eight fields. Coverage columns count expected means inside the corresponding pointwise 95\% intervals; fields with no intermediate blocks are excluded from that column. Last column: maximum absolute mean-area discrepancy in percentage points.}
\label{tab:conditional}
\begin{tabularx}{\linewidth}{@{}Xllrrrr@{}}
\toprule
Conditioned field & Base mesh & Format & Intermediate & Total CI & Mid CI & Max. diff.\\
\midrule
FP64 reference & 128 $\times$ 64 & BF16 & 12--55 & 7/8 & 7/8 & 0.239 \\
FP64 reference & 128 $\times$ 64 & FP16 & 6--95 & 8/8 & 8/8 & 0.222 \\
FP64 reference & 256 $\times$ 128 & FP16 & 28--236 & 8/8 & 8/8 & 0.090 \\
Sensor-only SR & 128 $\times$ 64 & BF16 & 12--59 & 8/8 & 8/8 & 0.136 \\
Sensor-only SR & 128 $\times$ 64 & FP16 & 6--97 & 7/8 & 7/8 & 0.486 \\
Sensor-only SR & 256 $\times$ 128 & FP16 & 28--235 & 8/8 & 8/8 & 0.094 \\
State + sensor SR & 128 $\times$ 64 & FP16 & 0--54 & 6/8 & 2/3 & 0.168 \\
\bottomrule
\end{tabularx}
\end{table}

\subsubsection{Effects of conserved-state rounding}\label{sec:state-sensor}

The following comparisons test the effect of conserved-state conversion on the pressure support described in Section~\ref{sec:pressure-method}.

We examine the undisturbed upstream normal-shock state ($\rho=P=1$, $u=1.1\sqrt{1.4}$, $v=0$) and a uniform downstream region of the coarse FP64 SV reference at $t=0.05$ (2,048 common-grid cells with identical conserved words). The upstream calculation uses the zero-vortex base state. The adjacent FP16 words of each conserved component are enumerated with probabilities from their exact decoded values; pressure is reconstructed using FP64 arithmetic. For the combined conversion, each reconstructed pressure is stochastically encoded and its probability mass is added to those of coincident words. Table~\ref{tab:pressure-support} compares the one-shot supports at $\tau=1.40625\times10^{-3}$. Direct pressure conversion is safe against uniform false tags in both regions. Rounding the conserved state produces a sufficiently large reconstructed-pressure range downstream, but the upstream range is still below threshold. Additional sensor conversion gives four upstream words spanning $2^{-9}$ across the binade boundary at one, with false-tag probability 0.28266. The two stages together can therefore create a false tag even where each individual conversion is acceptable.

\begin{table}[!tb]
\centering\small
\caption{One-shot independent FP16 conversion of the undisturbed upstream normal-shock state and the uniform downstream state. Pressure reconstruction uses FP64 operations while probabilities use the ideal adjacent-word model. The strict threshold is $1.40625\times10^{-3}$. Each tag probability concerns four distinct independently converted neighbors of an interior cell.}
\label{tab:pressure-support}
\begin{tabularx}{\linewidth}{@{}lXrrr@{}}
\toprule
Region & Conversion before comparison & Support size & Pressure diameter & False tags (\%)\\
\midrule
Upstream & Direct pressure only & 1 & $0$ & 0\\
 & Conserved words, then reconstruction & 4 & $1.289558\times10^{-3}$ & 0\\
 & Conserved words, reconstruction, sensor SR & 4 & $1.953125\times10^{-3}$ & 28.266\\
\addlinespace
Downstream & Direct pressure only & 2 & $9.765625\times10^{-4}$ & 0\\
 & Conserved words, then reconstruction & 8 & $1.457744\times10^{-3}$ & 3.350\\
 & Conserved words, reconstruction, sensor SR & 3 & $1.953125\times10^{-3}$ & 20.202\\
\bottomrule
\end{tabularx}
\end{table}

The coarse FP16 results illustrate this effect on the evolving mesh. At initialization, 176 of 5,394 interior downstream cells with identical unconverted pressure in their five-cell stencils tag from reconstructed pressure alone (3.263\%); after sensor conversion, 1,142 tag (21.172\%). The one-shot probabilities in Table~\ref{tab:pressure-support} explain the magnitude of these counts, with dependence between overlapping stencils. On the complete initial field, the reconstructed-pressure-only indicator selects 224 of 512 blocks and the combined indicator selects 502. Over the seven later regrids, reconstructed-pressure variation on the combined trajectory selects 509--512 blocks before additional sensor conversion, compared with 511--512 after conversion. 
The eight-seed combined FP16 SR ensemble refines 99.73\% of the domain on average, compared with 21.71\% for sensor-only FP16 SR. Thus sensor encoding is especially important at initialization, whereas the later mesh also responds to variation introduced by state storage and its subsequent propagation.

\subsubsection{Scientific observables and physical accuracy}\label{sec:physical}

The shock--vortex and scalar transport experiments of Section~\ref{sec:amr-experiments} relate the mesh and work changes to the error in physical quantities.

The usefulness of additional resolution depends on the physical quantity of interest. In coarse SV, the final maximum linewise shock-position errors relative to the twice-finest uniform reference are 0.002046 for BF16 RN and 0.002087 for the BF16 SR mean. The difference between these errors is substantially smaller than the variation of 0.002091 among the FP64 reference calculations. Similarly, the downstream RMS pressure perturbation varies by $5.23\times10^{-3}$ among the reference calculations. The sensitivity of these quantities to discretization prevents a conclusive assessment of whether the additional refinement improves physical accuracy.

\begin{figure}[!tb]
\centering
\includegraphics[width=\linewidth]{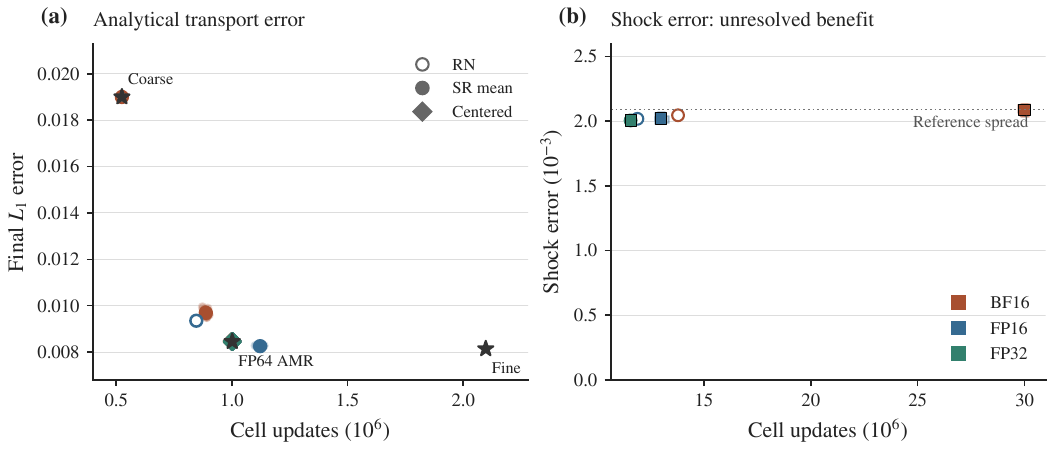}
\caption{\textbf{Numerical work must be interpreted with the scientific accuracy target.} (a) Smooth scalar transport with an analytical translating reference, base grid $64^2$: native and centered sensors, RN and SR seed means with individual SR realizations. Fixed-grid and FP64 adaptive controls are shown with stars. (b) Nominal SV on the $128\times64$ base grid, with shock-position errors against the twice-finest uniform reference. The dotted line shows the empirical reference spread; the observable does not resolve a benefit of the additional downstream refinement.}
\label{fig:accuracy}
\end{figure}

For smooth scalar transport, the analytical solution permits a direct comparison of error and work. The low-background sensor is $B_s+(g/2)q$, where $B_s=1.5$ is an even BF16 word, $g=2^{-7}$ and $0\le q\le1$. RN returns $B_s$ throughout the field, whereas SR chooses the upper word with probability $q/2$. This alignment produces amplitude-dependent dither. At base resolution $64^2$, the final $L_1$ error and work are 0.018998 and 524,288 for BF16 RN, compared with 0.009694 and 887,219 for the SR means. Direct fluctuation encoding with BF16 RN gives 0.008459 and 1,000,784 updates; uniform finest-grid evolution gives 0.008144 and 2,097,152 updates (Fig.~\ref{fig:accuracy}a).

Changing only the background to $B_s=1.5+\theta_0g$ alters its lattice phase, with the physical field, threshold, solver and seeds fixed. The sensor is converted from the exact dyadic affine sum of its decoded operands. In Table~\ref{tab:scalar-phase}, the off-lattice SR fields lose much of the adaptive work saving because nearly uniform regions can tag. The upper-word probability within the first interval is $\theta_0+q/2$, with the bracket updated when an interval boundary is crossed. At phase $3/4$, the sensor is a lattice word at $q=1/2$, reducing the probability of tagging near that contour. Across 32 seeds and eight regrids, the band $|q-1/2|\le0.1$ contains 3.47\% of cell observations but 48.97\% of unrefined observations. Its mean conditional tag probability is 0.143 versus 0.717 outside the band; the mean unrounded indicator is 0.0679 versus 0.0108. Thus a small under-resolved region near a large gradient can contribute to higher error even when most of the domain is refined.

At phase $1/2$, RN is sensitive to the arithmetic of sensor formation. The positive initial cell averages (minimum approximately $1.40\times10^{-23}$) put every exact affine sensor above the midpoint. Forming the sum in FP64 first absorbs the fluctuation in 124 cells, producing ties to the even lower word. This alternative gives 156 initial tags (3.81\% of base cells), 7.32\% initial refined area, 1.184\% update-weighted area and 542,912 updates, with the same final error 0.018998 at the displayed precision. Table~\ref{tab:scalar-phase} uses exact affine conversion throughout.

\begin{table}[!tb]
\centering\small
\caption{Scalar background-phase control on the $64^2$ base grid: $B_s=1.5+\theta_0g$, BF16 sensor with signal amplitude $g/2$. Physical inputs and the threshold are unchanged. Sensor conversion uses the exact affine sum of decoded operands; phase $1/2$ is sensitive to FP64 pre-addition (see text). SR entries are means of 32 runs at each phase. Work is in millions of cell updates. The uniformly refined comparator has $L_1=0.008144$ and work 2.097 million.}
\label{tab:scalar-phase}
\begin{tabular}{@{}lrrrrrr@{}}
\toprule
&\multicolumn{2}{c}{$L_1$ error}&\multicolumn{2}{c}{Refined area (\%)}&\multicolumn{2}{c}{Work}\\
$\theta_0$ & RN & SR & RN & SR & RN & SR\\
\midrule
0 & 0.018998 & 0.009694 & 0.00 & 23.07 & 0.524 & 0.887 \\
1/8 & 0.017958 & 0.008227 & 2.95 & 95.26 & 0.571 & 2.023 \\
1/4 & 0.015615 & 0.008147 & 5.36 & 99.82 & 0.609 & 2.094 \\
1/2 & 0.018998 & 0.008214 & 0.00 & 99.64 & 0.524 & 2.091 \\
3/4 & 0.018998 & 0.010380 & 0.00 & 97.25 & 0.524 & 2.054 \\
\bottomrule
\end{tabular}
\end{table}

\subsubsection{Rayleigh--Taylor instability: resolution-dependent refinement}\label{sec:rt-results}

The three Rayleigh--Taylor meshes in Section~\ref{sec:amr-experiments} test the one-bit-per-halving prediction of Eq.~\eqref{eq:spacing} using density instead of pressure as the sensor.

The native density threshold is $\tau=2/N_y$. For BF16, the spacing in $[1,2)$ is $2^{-7}$ and that in $[2,4)$ is $2^{-6}$. Thus, as the base vertical resolution increases from 128 to 256 and then 512, the threshold decreases from $2^{-6}$ to $2^{-7}$ and $2^{-8}$. At the coarsest resolution, neither density binade permits the uniform-field false-tag mechanism. At the intermediate resolution, the mechanism is possible in the higher binade only, principally for heavy-fluid densities slightly above 2. At the finest resolution it is possible in both binades. Exactly represented densities, including the ideal plateau values 1 and 2, have zero false-tag probability under uniform conditions; the mechanism acts on nonrepresentable values in the evolving density field. The spacing condition identifies the eligible ranges; the refinement depends on the local density distribution, nonuniform differences and block selection.

The selected meshes exhibit this progression. At $32\times128$, the FP64, BF16 RN and BF16 SR trajectories have identical refined areas at all 31 observations and identical density, marker and mesh arrays at the two saved field times; cumulative work through $t=3.75$ is also identical, although individual tags can differ. At $64\times256$, the refined fractions at $t=0.25$ are 9.375\%, 9.375\% and 22.656\%, respectively. The excess area in the SR mesh is confined to the upper half of the domain: its refined fraction there is 35.94\%, compared with 9.375\% for FP64, while the lower-half fractions are equal. At $t=3.75$, the FP64 and RN meshes each refine 20.90\% of the domain, whereas SR refines 56.74\%. At that observation, 98.86\% of the SR trajectory's base cells with density at least 2 lie in refined blocks, compared with 7.18\% in the FP64 trajectory, using each field's own conservatively averaged base-cell density. This is consistent with the higher-binade spacing crossing the tag threshold. At $128\times512$, the early refined fractions are 7.81\%, 10.11\% and 49.49\%, and by $t=3.75$ they reach 44.73\%, 58.01\% and 99.66\% (Fig.~\ref{fig:rt}); the SR trajectory uses 2.015 times the FP64 leaf-cell updates. Section~\ref{subsec:atlas-amr-application} reports the four-seed means. Unlike conditional draws on a frozen field, these comparisons allow the selected mesh to influence the subsequent state and thus test the consequences of the spacing mechanism in an evolving instability.

The physical solution remains sensitive to numerical resolution. The tracer interface develops a falling spike and roll-up on the finest base mesh, whereas the coarsest mesh has no interface overhangs at the late observation. The FP64 mixing width increases from 0.18083 at $64\times256$ to 0.42602 at $128\times512$. The mean density difference between those two FP64 solutions is 0.208531 on the common $128\times512$ comparison grid, compared with matched fine-grid differences of 0.018248 and 0.033471 for RN and SR relative to FP64. The grid-to-grid differences exceed the differences between rounding policies.

\FloatBarrier
\subsection{RQ3: Application of the Precision Atlas to adaptive mesh refinement}\label{subsec:atlas-amr-application}

\subsubsection{Atlas-guided refinement requirements}

The Atlas entry in Section~\ref{sec:atlas-tag-entry} is first used to determine the sensor precision required on the Rayleigh--Taylor grids.

The Atlas provides a lookup of representation requirements. Its pressure-difference entries G1/G2 (Table~\ref{tab:matched}) relate sensor magnitude and spacing to resolvable variations; the energy entries E1/E2 describe the effect of the representation used to obtain pressure. For density the same spacing argument applies. On the $128\times512$ RT grid, $\tau=2^{-8}$, so Eq.~\eqref{eq:spacing} requires at least nine significand bits in $[1,2)$ and ten in $[2,4)$ to exclude uniform-field SR tags. BF16 supplies eight. The Atlas thus identifies density ranges for which refinement can be triggered by rounding alone.

\subsubsection{Refinement under reduced-precision sensor storage}

The sensor-only experiments compare direct tagging with the Atlas-guided criterion of Section~\ref{sec:guided-method} on both shock--vortex meshes and all three Rayleigh--Taylor meshes.

\begin{table}[!ht]
\centering\footnotesize
\caption{Direct and Atlas-guided tagging on the two SV grids and three late-time RT grids. Each entry is RN / SR. Extra tags are candidate-only decisions as a percentage of all evaluated decisions; retained tags are the run-mean percentage of reference-positive decisions tagged by the guided rule. Area is the update-weighted refined fraction (\%), derived from $W=N_0K(1+3\overline A)$ for the timestep schedule in Section~\ref{sec:amr-experiments}. Thus area and work are not independent measurements. $W/W_{64}$ uses the matched FP64 reference, unlike the SR/RN ratio in Table~\ref{tab:work}. SR uses 16 seeds for SV and four for RT. All guided extra-tag counts are exactly zero.}\label{tab:atlas-amr-application-tags}
\setlength{\tabcolsep}{2.5pt}
\begin{tabular}{@{}llrrrrrrr@{}}
\toprule
 & & \multicolumn{2}{c}{Extra tags (\%)} & Retained & \multicolumn{2}{c}{Area (\%)} & \multicolumn{2}{c}{$W/W_{64}$}\\
\cmidrule(lr){3-4}\cmidrule(lr){6-7}\cmidrule(l){8-9}
Case/grid & Format & Direct & Guided & Guided (\%) & Direct & Guided & Direct & Guided\\
\midrule
SV $128\times64$ & BF16 & 3.45 / 63.50 & 0 / 0 & 45.30 / 37.49 & 25.1 / 93.8 & 8.9 / 7.8 & 1.19 / 2.59 & 0.86 / 0.84 \\
SV $128\times64$ & FP16 & 0.50 / 1.50 & 0 / 0 & 81.79 / 72.13 & 17.1 / 21.7 & 14.0 / 12.7 & 1.03 / 1.12 & 0.97 / 0.94 \\
SV $128\times64$ & FP32 & 0 / 0 & 0 / 0 & 100.000 / 100.000 & 15.7 / 15.7 & 15.7 / 15.7 & 1.00 / 1.00 & 1.00 / 1.00 \\
SV $256\times128$ & BF16 & 1.61 / 62.81 & 0 / 0 & 26.13 / 20.03 & 16.6 / 91.3 & 4.9 / 4.2 & 1.08 / 2.69 & 0.82 / 0.81 \\
SV $256\times128$ & FP16 & 7.93 / 41.73 & 0 / 0 & 75.86 / 61.67 & 29.7 / 86.8 & 10.5 / 9.0 & 1.36 / 2.59 & 0.95 / 0.91 \\
SV $256\times128$ & FP32 & 0 / $<0.001$ & 0 / 0 & 99.996 / 99.990 & 13.0 / 13.0 & 13.0 / 13.0 & 1.00 / 1.00 & 1.00 / 1.00 \\
RT $32\times128$ & BF16 & 0 / 0.008 & 0 / 0 & 80.10 / 79.75 & 12.5 / 12.5 & 12.5 / 12.5 & 1.00 / 1.00 & 1.00 / 1.00 \\
RT $64\times256$ & BF16 & 0.71 / 10.74 & 0 / 0 & 73.37 / 66.91 & 13.1 / 35.3 & 7.7 / 6.8 & 1.03 / 1.52 & 0.91 / 0.89 \\
RT $128\times512$ & BF16 & 3.15 / 31.71 & 0 / 0 & 70.00 / 48.60 & 27.7 / 71.8 & 13.7 / 10.7 & 1.17 / 2.01 & 0.90 / 0.84 \\
\bottomrule
\end{tabular}
\end{table}

\begin{figure}[!ht]
\centering
\includegraphics[width=.92\linewidth]{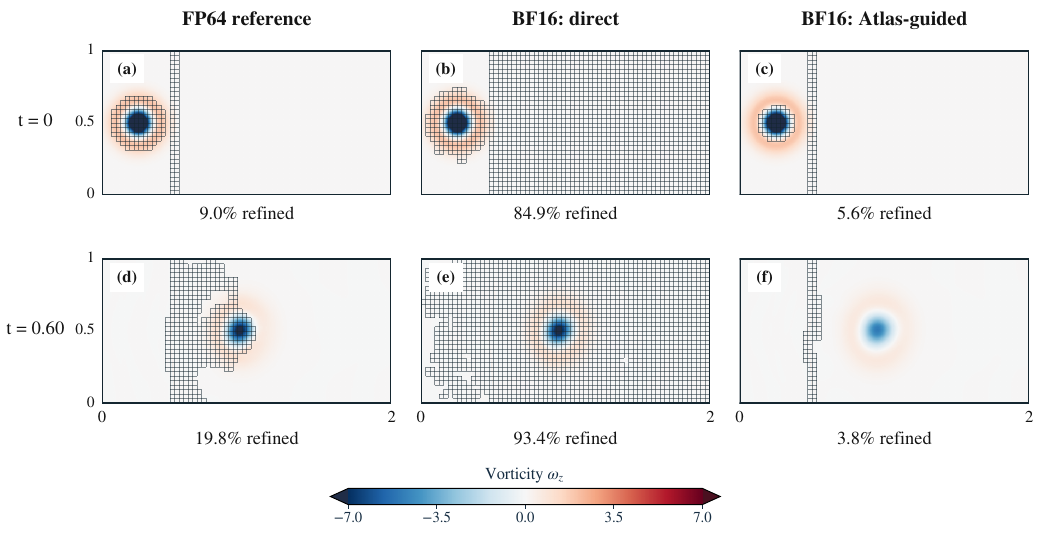}
\caption{Vorticity and active blocks in SV on the $256\times128$ base grid: FP64 reference, direct BF16 SR and Atlas-guided BF16 SR (seed zero). Initial panels show the mesh after the first remap; final panels show the active mesh before remapping. Percentages denote refined area.}\label{fig:atlas-amr-sv}
\end{figure}

On the finer SV grid, direct BF16 SR tags much of the downstream fluid, whereas the guided criterion excludes those tags but loses refinement around the vortex (Fig.~\ref{fig:atlas-amr-sv}). At $t=0.60$, the refined fractions in the illustrated realization are 19.8\% for FP64 and 3.8\% for guided BF16. Table~\ref{tab:atlas-amr-application-tags} shows that the mean retained fraction under SR increases from 20.03\% (BF16) to 61.67\% (FP16) and 99.99\% (FP32). The guided rule is a diagnostic of insufficient precision when the retained fraction is low. The direct BF16 SR work ratios on the coarse and fine grids are 2.59 and 2.69 relative to FP64 (Table~\ref{tab:atlas-amr-application-tags}), or 2.17 and 2.50 relative to the corresponding BF16 RN calculations (Table~\ref{tab:work}). These describe the same work counts with different denominators.

\begin{figure}[!ht]
\centering
\includegraphics[width=.85\linewidth]{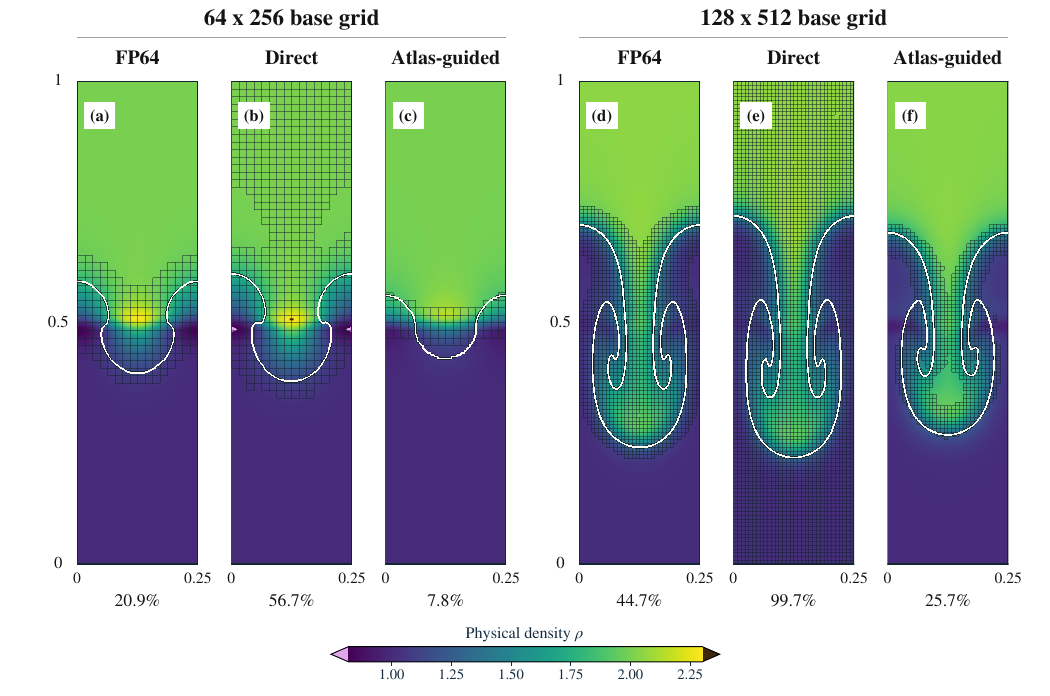}
\caption{Density, active blocks and the $C=0.5$ interface at $t=3.75$ on the $64\times256$ and $128\times512$ RT grids. Direct and Atlas-guided BF16 SR use seed zero. Percentages denote refined area. At the intermediate resolution, the guided rule removes the overhangs; at the fine resolution, roll-up remains visible.}\label{fig:atlas-amr-rt}
\end{figure}

For RT at $t=3.75$, the four-seed mean refined areas on the finest grid are 44.73\% (FP64), 99.82\% (direct BF16 SR) and 25.51\% (guided BF16 SR). The interface roll-up remains visible (Fig.~\ref{fig:atlas-amr-rt}), but on the intermediate grid the guided rule eliminates overhangs in all four seeds, compared with 16 of 128 overhang columns in the FP64 reference. For $64\times64$ scalar transport with native BF16 SR, all decisions are unresolved. Over 32 seeds, the final analytical $L_1$ errors are 0.00846 (FP64), 0.00969 (direct BF16) and 0.01900 (guided BF16, unrefined). Reducing refinement is therefore not itself evidence of improved accuracy.

When a wider sensor already exists, the unresolved comparison can instead be evaluated with that sensor, as in adaptive-precision filters~\cite{Shewchuk1997}. The separate coarse native BF16 SR control reproduces the reference decisions but recomputes approximately 96.4\% of cell decisions. The choice of leaving unresolved cells untagged, using wider inputs where available or changing the representation depends on the required physical accuracy and mesh.

\subsubsection{Combined state and sensor storage}

The combined-state experiments of Section~\ref{sec:amr-experiments} compare a sensor-spacing radius with the composed bound of Section~\ref{sec:guided-method}.

In a paired comparison on nine states (one RN and two SR realizations, each at $t=0,0.3,0.6$), the sensor-only radius fails to enclose the pre-conversion pressure in 5,222 of 73,728 cells (7.1\%). It produces six extra tags (0.0081\% of all decisions) and misses 17,425 of the 28,146 reference-positive decisions. The composed radius encloses all reference pressures and produces no extra tags, but misses 22,918 reference-positive decisions, an increase of 5,493. The benefit in excluding extra tags in these fields is small; the pressure reconstruction term is required for the stated enclosure, rather than evidence of a large accuracy benefit. Across the evolved combined-state calculations (one RN and eight SR realizations), the composed criterion leaves 87.44\% (RN) and 94.73\% (SR) of all decisions unresolved, and retains 58.75\% and 9.31\% of reference-positive decisions, respectively. Thus FP16 state and sensor storage at the specified threshold rarely permits a resolved comparison under this bound.\par
For SR, the mean update-weighted refined area is 88.91\% for the pre-conversion reference, versus 15.72\% for the all-FP64 adaptive calculation. The disparity is consistent with the refinement induced by prior state rounding; retention relative to this pre-conversion reference is not retention of the all-FP64 tags. The guided calculation has an update-weighted refined area of 11.94\% and uses 0.370 times the pre-conversion-reference leaf work, or 0.923 times the all-FP64 leaf work (mean updates: $1.07\times10^7$, $2.88\times10^7$ and $1.16\times10^7$ for the guided, pre-conversion-reference and all-FP64 cases, respectively). The corresponding mean final pressure $L_1$ errors are $2.127\times10^{-3}$, $1.814\times10^{-3}$ and $1.418\times10^{-3}$, against the same all-FP64 uniform grid with four times the base resolution in each direction. The guided policy reduces extensive refinement at the expense of a larger pressure error; it uses about 8\% fewer leaf updates than all-FP64 but has about 50\% larger pressure error. 

\FloatBarrier

\section{Discussion}\label{sec:discussion}

The Atlas connects the arithmetic requirements of the sensor-producing operation to those of the adaptive discretization. The spacing condition predicts where uniform-field SR can introduce extra tags; pressure reconstruction from rounded conserved quantities can broaden the pressure distribution even when direct sensor conversion satisfies that condition. The observed RT progression from heavy-fluid refinement to nearly domain-wide refinement illustrates the spatial nature of this limit. The joint block probabilities translate local sensor distributions into expected mesh size and, for the specified timestep policy, cell-update work. This complements discretization-error-based precision selection and precision assignment across AMR levels~\cite{RAPTOR2025,Mazumder2026,Tamstorf2021}.

The bound-based criterion in Section~\ref{subsec:atlas-amr-application} uses the same error-bound principle as adaptive-precision predicates~\cite{Shewchuk1997,Titolo2019}. Its purpose is to use the Atlas bounds for the particular physical quantities in the sensor, not to introduce a new predicate algorithm. The combined-state test demonstrates why sensor spacing alone is insufficient to enclose the pre-conversion pressure: reconstruction propagates the errors of density, momentum and energy into the tag decision. The six extra tags removed by the composed radius on the paired fields are few compared with the additional missed reference tags. The large unresolved fractions in the evolved FP16 runs are the principal indication of insufficient representation for this criterion. The local uncertainty interval distinguishes a resolved above-threshold variation from a resolved below-threshold variation; it does not classify unresolved variations by physical importance. Leaving them untagged can suppress required refinement, as the BF16 scalar, SV and intermediate-grid RT results show. The retained fraction of reference tags, solution error and mesh geometry must therefore be considered together. The same principle applies to other threshold-dependent operations in the Atlas, such as HLL flux branches and phase or root decisions.

Alternative sensor representations can avoid some of this sensitivity. In the coarse BF16 SV calculation, common-variate SR reduces mean extra tags from 63.50\% to 3.32\%; evaluating the difference in FP64 before converting it reduces the fraction to approximately $0.0002\%$. Direct offset storage reduces the extra-tag fraction of SR to 1.34\% on the coarse grid, but gives 38.71\% on the finer grid. These comparisons and the higher-precision comparison in Section~\ref{subsec:atlas-amr-application} illustrate distinct choices of representation and arithmetic order. Conserved-state conversion must also preserve the physical relationships between components, such as $0\le\rho C\le\rho$ in RT (Appendix~\ref{app:cfd}). The scalar phase comparison shows that additional refinement can improve the physical solution even when the tags differ from the reference policy. 

\paragraph{Limitations:} The calculations use a first-order scheme, two refinement levels and aligned blocks, with a common timestep and no subcycling. Sensor-only runs retain FP64 flow evolution; state-storage tests also retain FP64 flux and update arithmetic. Tag agreement is assessed against the unrounded sensor on the current trajectory. Work counts active leaf-cell updates and omits mesh generation, load balancing, data movement and hardware-dependent costs. Other refinement indicators, and lower-precision flux or stage arithmetic were not tested. RN comparisons are deterministic; the SR ensembles contain 16 seeds for SV and four for the RT tagging comparisons. 

\section{Conclusions}\label{sec:conclusion}

The precision required by numerical operations in scientific calculations was examined and applied to adaptive mesh refinement. The Precision Atlas organizes these requirements by the quantities represented, their physical scales, the arithmetic and the desired accuracy. The time integration controls show that an admissible advancing timestep does not ensure that the stored state or the output can resolve the associated change. In explicit diffusion, halving the mesh spacing increases the time-significand requirement by two bits at a fixed normal time. All eleven configurations selected before integration using sufficient bounds meet their accuracy targets.

In AMR, the quantities that determine the discretization introduce a further representation requirement. For a fixed gradient threshold, excluding phantom gradients (uniform-field false tags) under independent SR requires one additional significand bit for each halving of the mesh spacing at a fixed sensor binade. The shared-stencil probability model relates this condition to block selection and cell-update work. The calculations are consistent with this condition. In the SV problem, direct BF16 SR tagging requires 2.59 and 2.69 times the FP64 leaf-cell work on the coarse and fine grids, or 2.17 and 2.50 times the corresponding BF16 RN work. On the three RT grids, the extra BF16 refinement appears first in the heavy fluid and then across both density ranges, as the threshold falls below the spacing of each binade. The pressure reconstruction calculation further shows that conversion of the conserved state can violate a tagging requirement even when the pressure representation alone is adequate.

The Atlas bounds were then used in the tagging criterion. In all sensor-only comparisons in Table~\ref{tab:atlas-amr-application-tags}, the guided criterion gives no extra tags relative to the reference. On the finer SV grid, it retains 20.03\%, 61.67\% and 99.99\% of the reference tags under SR for BF16, FP16 and FP32, respectively. The tags lost at low precision affect the physical solution. Guided BF16 tagging removes the vortex refinement, leaves the scalar transport problem unrefined and suppresses the RT overhangs on the intermediate grid. When the conserved state and the sensor are both stored in FP16, the pressure reconstruction bound is required to enclose the pre-conversion pressure. On the paired fields, it removes six extra tags and adds 5,493 missed reference tags. In the evolved calculations, the composed criterion leaves 87.44\% (RN) and 94.73\% (SR) of the decisions unresolved. Under SR, the leaf-cell work is 0.370 of that of the extensively refined pre-conversion reference and 0.923 of that of the all-FP64 calculation, and the final pressure error is about 50\% larger than in the all-FP64 calculation. The criterion therefore determines whether the stored quantities resolve the requested refinement threshold. It does not ensure physical accuracy and cannot recover the information discarded by rounding.

These results show how the Atlas can inform the numerical design of adaptive calculations. The appropriate balance between precision, refinement and cost depends on the accuracy required of the application. The analysis can be extended to multilevel hierarchies, higher-order schemes and other threshold-dependent operations, such as flux and phase selection.

\section*{Acknowledgments}
This work was supported by the Center for Prediction, Reasoning, and Intelligence for Multiphysics Exploration (C-PRIME), a PSAAP-IV project funded by the U.S. Department of Energy under grant DE-NA0004264 (program manager: Dr. David Etim).

\section*{Data and code availability}
Selected numerical data and code will be provided as electronic supplementary materials accompanying the manuscript. Additional data are available from the authors upon reasonable request. 

\FloatBarrier
\appendix
\Needspace{8\baselineskip}
\section{Representation requirements for controlled kernels}\label{app:matched}

\subsection{Scope and exact width arguments}
For a bounded real-vector target with unrestricted output in the maximum norm, the optimal worst-case error on an input-encoding fiber is half the largest coordinate range on that fiber. Choosing the midpoint of each coordinate's infimum and supremum attains this radius. With restricted outputs the intersection of all acceptable output sets must additionally contain an allowed word. This standard radius-of-information argument supplies the lower and upper representation bounds used here; no nesting of input fibers across widths is assumed. All seven controlled problems use nearest-even binary arithmetic with eight exponent bits, gradual underflow, and candidate significand widths 4 through 53. The tolerance is $\varepsilon=2^{-q}$. Exact decoder arithmetic and candidate-format output are distinguished.

\paragraph{Pressure gradient (G1, G2).} Let $a_L,a_R\in[-1/8,1/8]$ independently, $h_P=1/32$, and $J_\nabla=(a_R-a_L)/(2h_P)$. For G1, store $B_P+a_i$ with known $B_P=3\,2^{e_P-1}$ and $e_P=0,4,\ldots,32$. The inputs remain in one binade. Their spacing is $g=2^{e_P+1-p}$ and exact differencing gives the bound $g/(2h_P)=2^{e_P+5-p}$. Swapping $B_P\pm\delta$ strictly inside a common cell gives target separation $2\delta/h_P$; one can choose $h_P\varepsilon<\delta<\min(1/8,g/2)$ at each excluded width. The resulting width is $e_P+5+q$, $q\in\{10,14\}$. The finite candidate subtraction is exact by Sterbenz's condition and power-of-two scaling preserves the word, so the same width suffices for that output. G2 supplies the fluctuations upstream. The maximum input rounding error is $2^{-p-4}$ and the gradient bound is $2^{1-p}$. A swap about $3/32$ with separation parameter $h_P\varepsilon<\delta<\min(1/32,2^{-p-4})$ supplies the lower witness. The unrestricted-output width is $q+1$, independent of $e_P$.

\paragraph{Gas pressure (E1, E2).} Set density to 2, momentum to $2^{k/2+1}$, $\gamma=3/2$, $K=2^k$ and internal energy $U\in[1,3]$. Here $k=4,8,\ldots,36$ and $q\in\{8,12\}$. The kinetic-energy calculation is exact at the stated powers of two. For stored total energy $K+U$, its rounding contributes at most $g/4=2^{k-p-1}$ to gas pressure $U/2$. At the sufficient widths the subtraction is exact by Sterbenz's condition and the pressure scaling is exact and positive. If $g\le2$, use colliding energies centered at $K+2$, with $2\varepsilon<\delta<\min(1,g/2)$. If $g\ge4$, $U=1$ and $3/2$ both encode at $K$ and give incompatible targets. This yields $k+q-1$. Supplying $U$ instead gives pressure error at most $2^{-p}$ and a matching witness about $U=5/2$, so the E2 width is $q$.

\paragraph{Linear systems (L1--L3).} Use $\lambda=2^{-k}$, $k=0,4,\ldots,32$, and the known exact matrices $A_D=\operatorname{diag}(1,\lambda)$ and $A_R=\tfrac12\begin{pmatrix}1+\lambda&1-\lambda\\1-\lambda&1+\lambda\end{pmatrix}$. The exact solution has components $3/2\pm d$ in the difference-direction cases, or both $3/2+d$ in the common-direction case, with $|d|\le1/8$. The input is the encoded right-hand side, with exact downstream decoding and unrestricted output. In L1 the power-of-two scaling of the second coordinate preserves its relative rounding grid, giving absolute error at most $2^{-p}$. In L2 the difference direction is amplified by $1/\lambda$, and $\|A_R^{-1}\|_\infty=1/\lambda$ gives $2^{k-p}$. The two probes $d=\pm\delta$ collide in the cell of $3/2$ when $\lambda\delta<2^{-p}$; choose $\delta>\varepsilon$ within the domain at every excluded width. In L3 the exact inverse preserves common vectors, as does identical RN of both input components, giving $2^{-p}$. The matching collision for L1/L3 uses $\delta<2^{-p}$. Thus the widths are $q,k+q,q$, respectively. All matrices have the same condition number $2^k$ at a given scale; the results instantiate effective conditioning~\cite{ChanFoulser1988}.

\subsection{A restricted-output witness for the centered gradient}
The G2 decoder can be required to return a candidate-format word, rather than an unrestricted exact real output. At $p=q+1$, set $g=2^{-p-3}$, $\delta=g/4$, $\ell=3/32$ and $r=-3/32-g$. The two inputs $(\ell+\delta,r-\delta)$ and $(\ell-\delta,r+\delta)$ both encode as $(\ell,r)$ and lie strictly inside the input domain and rounding cells. Their targets are $-3-3\varepsilon/2$ and $-3-\varepsilon/2$, so their common acceptable interval is $[-3-3\varepsilon/2,-3-\varepsilon/2]$. The adjacent candidate outputs $-3-2\varepsilon$ and $-3$ lie strictly outside this interval. Thus $q+1$ is excluded even though the targets differ by only $\varepsilon$. At $p\ge q+2$, the input error is at most $2^{1-p}$ after differencing and scaling, and candidate subtraction contributes at most another $2^{1-p}$; multiplication by 16 is exact within the supported range. The total is $2^{2-p}\le\varepsilon$, giving the exact first and sustained width $q+2$. This is a separate output requirement, with two distinct tolerance problems repeated across the nine irrelevant background settings. 

\section{Time and accumulation bounds}\label{app:time}
The technical supplementary material derives the rounding-cell endpoints, arithmetic sequences, normality conditions and reference-resolution checks used in the time integration examples.

For a stored time $z$ with a finite successor, an advancing step exists when $\operatorname{succ}(z)-z\le H$. Directed selection then returns the largest stored time not exceeding $z+H$, whereas a nearest-rounded request must also satisfy the midpoint and tie conditions given in the Supplementary Material. The local Robertson ceiling for explicit Euler is $\min_{\lambda\ne0}[-2\operatorname{Re}\lambda/|\lambda|^2]$, evaluated over the modes of the frozen Jacobian with nonpositive real part. For the scalar comparison $v'=-\mu v$, $\mu=250001/25$, the Euler and classical RK4 stability polynomials require 50 and 49 time-significand bits at $T=10^{11}$, respectively. Requiring an exactly represented RK4 midpoint raises its requirement to 50 bits.

For the periodic diffusion tests selected before integration ($N\in\{48,96\}$, $\kappa=3/2$, $L=2^{-9}$), the error bound is the sum of arithmetic, temporal, spatial and initialization terms. The Supplementary Material derives these terms from the nonexpansivity of the Euler update under its step ceiling, the accumulation of local residuals and consistency with the continuous solution. For $K$ updates and FP64 unit roundoff $u_{64}=2^{-53}$, the bound requires $K\le256$, $10Ku_{64}<1$, normal durations within the stability ceiling, and exact attainment of the four prescribed output times. For the uniform source with initial value $a$, the exact trajectory is $a+\sigma\xi$, and the FP32 products are accumulated in a single FP32 word, a compensated FP32 pair or an FP64 accumulator. The Supplementary Material bounds the formation of the products and their accumulation separately, for at most 128 updates and under stated conditions on the lattice and magnitude of the values, and gives a separate bound for the output. The 427-update history in Fig.~\ref{fig:source} exceeds this limit and is assessed by exact arithmetic comparison.

For the four diffusion and eight uniform-source cases described in Section~\ref{sec:time-results}, the pre-integration procedure selects eleven configurations, and all eleven meet their accuracy targets. The remaining case permits only FP32 output at tolerance $2^{-30}$, and 88 of its prescribed output values have no acceptable FP32 word. Five other comparison configurations meet their targets although their sufficient bounds exceed the tolerances. At tolerance $2^{-30}$, the source output requires 30 bits over both histories and 18 bits at the final time. The Supplementary Material verifies the arithmetic bounds using exact residuals and rounding identities; the Robertson comparisons rely on numerical references whose resolution is checked empirically.

\FloatBarrier

\section{Additional integration and linear-solve controls}\label{app:rk-cg}
\subsection{A matched Runge--Kutta lattice experiment}
\begin{figure}[!ht]
\centering
\includegraphics[width=.88\linewidth]{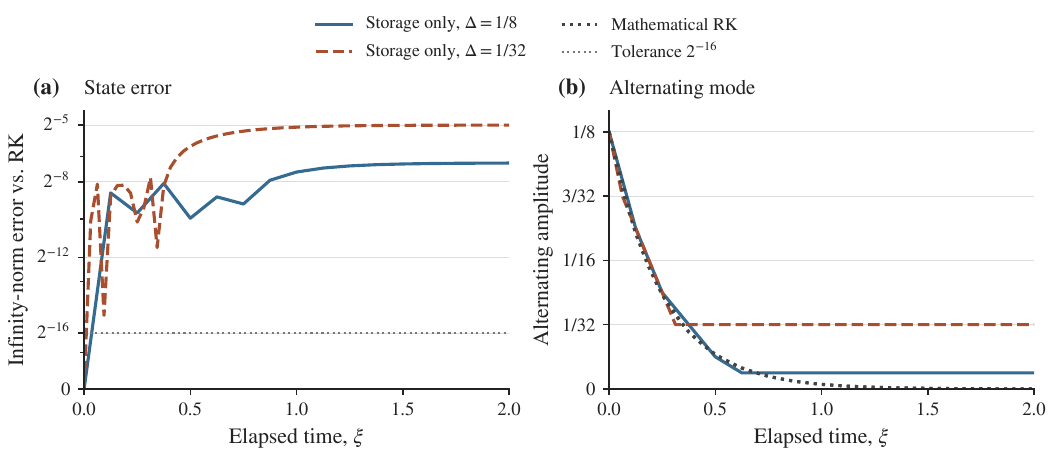}
\caption{RK3 storage-only controls: exact discrete reference, represented states and repeated-state-rounding errors on the matched linear problem. The analytic lattice recurrence explains the nonzero stationary amplitude for the two timesteps $\Delta$.}
\label{fig:rkstorage}
\end{figure}

Figure~\ref{fig:rkstorage} records the exact recurrence of a single represented Fourier mode, which provides a small benchmark for the stored-state mechanism analyzed more generally by Barad~\cite{Barad2026}. On an eight-node periodic grid, apply the discrete diffusion operator with alternating-mode eigenvalue $-4$. The initial state is $B+(-1)^j/8$, with $B=3/2$ or 98304; the tested exact RK3 timesteps are $\Delta=1/8,1/32$ up to elapsed time 2. The classical Kutta and third-order SSP formulas have the same linear exact step map $R(-4\Delta)$, $R(z)=1+z+z^2/2+z^3/6$. However, their internal arithmetic differs as noted in Ref.~\cite{Ketcheson2014}. The storage-only arm evaluates that map exactly and then stores each state. At $B=98304$ and $p=24$, write the represented alternating amplitude as $m_n g$, $g=1/128$. The background has even lattice parity. The amplitude satisfies $m_{n+1}=\operatorname{RN}_{\mathbb Z}(r m_n)$ with $r=29/48$ or $2711/3072$. The rounding cells produce nonzero stationary amplitudes $g$ and $4g$, respectively; the first completely absorbed steps are 6 and 11. The smaller timestep lowers the exact RK discretization error but increases the repeated-storage error. The recurrence provides an exact discrete example of the storage-induced turnaround analyzed by Barad.

\begin{figure}[!ht]
\centering
\includegraphics[width=.78\linewidth]{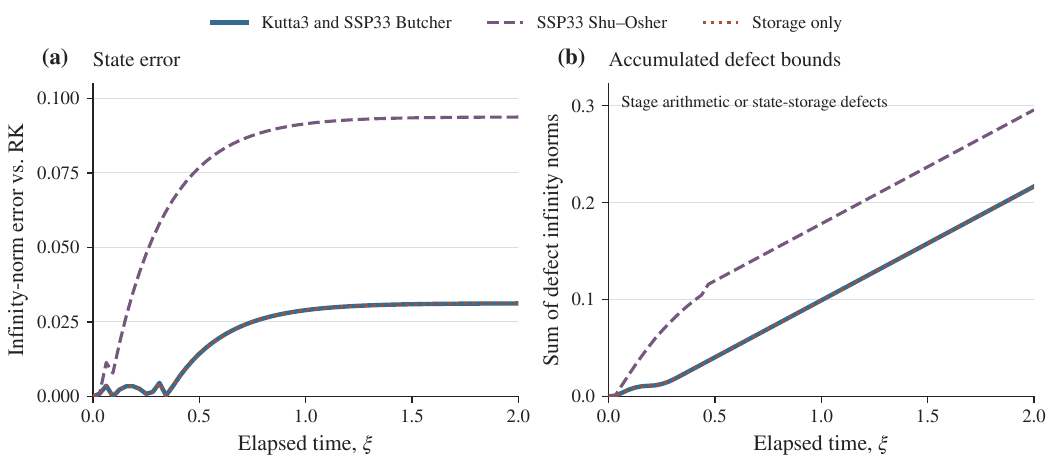}
\caption{Internal-arithmetic RK comparison for algebraically matched exact linear step maps.}
\label{fig:rkinternal}
\end{figure}

The internal-rounding arm (Fig.~\ref{fig:rkinternal}) uses the same exact step map but candidate arithmetic at the stages. The rounded Shu--Osher implementation has a mean drift of $1/16$ and maximum exact-map error 0.093708 for the large-background FP32, $\Delta=1/32$ case. The distinct Kutta/Butcher and Shu--Osher paths illustrate internal error propagation and rounding of affine combinations. For repeated single-word storage of the exact map, a uniform local defect bound $g/2$ and $0<r<1$ give $|e_n|\le(g/2)(1-r^n)/(1-r)$ when the initial state is exact. The first-step output-distance exclusions and this geometric sufficient bound give width brackets $[32,34]$ and $[32,36]$ for the two large-background timesteps at tolerance $2^{-16}$, and $[16,18]$, $[16,20]$ for the smaller background. These bracket first feasible widths under the specified map and candidate family; the sufficient tail bound is monotone, but the finite trajectory is not assumed monotone.

\subsection{Conjugate-gradient Atlas entry}
\begin{figure}[!ht]
\centering
\includegraphics[width=.68\linewidth]{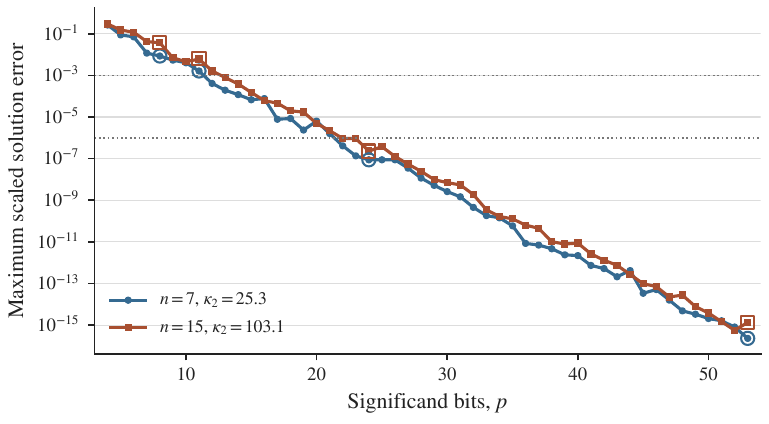}
\caption{Unpreconditioned conjugate-gradient measurements across the custom significand family and named formats. The errors are recorded for two fixed problems.}
\label{fig:cg}
\end{figure}

Unpreconditioned conjugate gradient (CG) is evaluated for two fixed systems of dimensions 7 and 15 (condition numbers approximately 25.27 and 103.1). At scaled tolerance $10^{-3}$ the sustained observed widths are 12 and 13; at $10^{-6}$ both are 22 (Fig.~\ref{fig:cg}). In Fig.~\ref{fig:cg}, the maximum scaled solution error is the relative infinity-norm error of the solution obtained using CG at a given precision and is defined as $\|\boldsymbol{x}_p-\boldsymbol{x}_{\mathrm{ref}}\|_\infty/ \|\boldsymbol{x}_{\mathrm{ref}}\|_\infty$, where the reference is obtained using an FP64 direct solve. This error metric is used for all 60 kernels in the Atlas.

\FloatBarrier

\section{Supporting threshold and physical-quantity controls}\label{app:decisions}
\subsection{An exact nonmonotone root decision}
\begin{figure}[!ht]
\centering
\includegraphics[width=.91\linewidth]{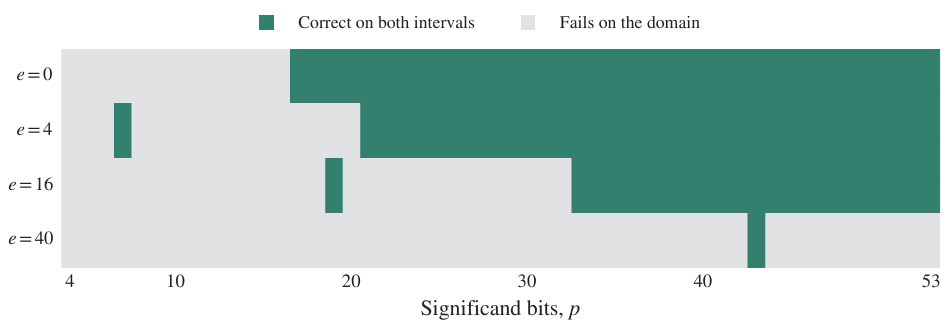}
\caption{Analytically determined feasible widths of the manufactured root predicate. The isolated passing point occurs when a rounding boundary aligns with the physical threshold.}
\label{fig:nonmonotone}
\end{figure}

The manufactured AMR root predicate stores $B$ and $B+s$, where $B=3\,2^{e-1}$, $\tau=1/8$ and $s\in[\tau-2m,\tau-m]\cup[\tau+m,\tau+2m]$, $m=2^{-16}$. The reference tags when $s>\tau$. In the tested normal range the specified RN implementation has the exact feasible set
\begin{equation}
\mathcal F_{\rm root}=\big(\{e+3\}\cup[e+17,\infty)\big)\cap\mathcal P.
\end{equation}
For $p\le e+2$ the gap is at least $1/2$ and both sides round to the background. At $p=e+3$ the gap is $1/4$ and the cell boundary coincides with $\tau$: the two sides round to 0 and $1/4$ relative to the background. For $e+4\le p\le e+16$, $\tau$ is an integer multiple of the gap and the upper inner margin can round to $\tau$, with the midpoint at $p=e+16$ resolved to the even word. The strict predicate then misses a required tag. For $p\ge e+17$ the spacing is at most $m$, and the error cannot change the decision. Figure~\ref{fig:nonmonotone} distinguishes the isolated feasible width from the sustained interval; $e=4$ is an analytic illustration, and at $e=40$ the sustained tail is outside $\mathcal P=\{4,\ldots,53\}$. This is a property of the specified root predicate, and is not a lower bound for arbitrary decoders or the entire adaptive tree.

The quadratic-profile example also distinguishes tag reproduction from physical accuracy: the one-cell projection error $2s/3$ exceeds $1/16$ even at the lower margin, while the two-cell projection has error $5s/12$ below tolerance.

\subsection{Branch labels and physical quantities}
In the HLL near-branch example the chosen sign margin $2^{-16}$ gives a 16-bit branch requirement. The flux changes by at most $2^{-13}$ from its boundary value, so a constant boundary flux already meets the $2^{-12}$ tolerance at the search floor $p=4$. In the fixed-$K$ flash and symmetric regular-solution phase models the label and composition gap have different precision requirements from the thermodynamic energy. At $\chi=2+2^{-20}$, the reference gap is about $1.196\times10^{-3}$ while the free-energy reduction is about $1.705\times10^{-13}$. For the near-equilibrium kinetics example the target is the net rate $d(3+d)$. Each accuracy specification distinguishes the quantity of interest from the numerical decision producing it.

\FloatBarrier

\section{Adaptive CFD setup and secondary comparisons}\label{app:cfd}
For RT, let $U=(\rho,\rho u,\rho v,E)$ and let $B$ denote the initialized hydrostatic cell-average state with zero velocity. The flux for these four conserved components is
\[
\widehat F^\Delta=\widehat F_{\rm HLL}(U_L,U_R)
-\widehat F_{\rm HLL}(B_L,B_R),
\qquad S^\Delta=(0,0,-(\rho-\rho_B),-\rho v).
\]
Here $S^\Delta$ is the remaining gravitational source after cancellation of the equilibrium pressure force and gravity. The unperturbed hydrostatic state $U=B$ therefore has zero update. The marker flux is $\widehat F^\Delta_\rho C_{\rm donor}$, where the sign of the density flux selects the upwind marker fraction. For SV, the same rule uses the HLL mass flux without equilibrium subtraction.

For RT, refinement adds the parent perturbation $U_p-B_p$ to each child's hydrostatic state. Marker mass is prolonged in the same way; its four child deviations from $\rho_jC_p$, where $C_p$ is the parent marker fraction, are scaled by the largest common factor in $[0,1]$ that enforces $0\le\rho_jC_j\le\rho_j$ while preserving the parent marker integral. SV uses piecewise-constant prolongation. Covered parent states are obtained by averaging their four equal-area children. At outflow boundaries the adjacent interior state is copied; at reflecting boundaries its normal momentum is reversed. The inflow state is prescribed below. The strict tag test, one-face-neighbor buffer and aligned $4\times4$ parent blocks determine the active mesh. Observation fields use the active mesh immediately before remapping.

The RT domain is $[0,1/4]\times[0,1]$, with gravity $(0,-1)$, periodic $x$ boundaries and reflecting $y$ boundaries. With $d=1/64$, the initial density is $\rho_0(y)=3/2+(1/2)\tanh((y-1/2)/d)$, and the hydrostatic pressure satisfies $dP_0/dy=-\rho_0$, $P_0(0)=3$. The divergence-free velocity perturbation is
\[
u_0(x,y)=-\frac{A\pi}{k}\sin(kx)\sin(2\pi y),
\qquad
v_0(x,y)=A\cos(kx)\sin^2(\pi y),
\qquad k=8\pi.
\]
The nominal, weak and zero perturbations use $A=0.01$, $0.001$ and $0$, respectively. The initial marker fraction is $C_0(y)=\rho_0(y)-1$.

\Needspace{9\baselineskip}
For SV, the domain is $[0,2]\times[0,1]$, and the shock is at $x=1/2$. Let $M_1=1.1$, $U_\infty=M_1\sqrt{\gamma}$, $X=(x-1/4)/r_c$, $Y=(y-1/2)/r_c$, $r_c=0.05$, and $G=\exp[\alpha(1-X^2-Y^2)]$ with $\alpha=0.204$. Upstream of the shock, the isentropic vortex has
\begin{align*}
u_0&=U_\infty+\epsilon YG, & v_0&=-\epsilon XG,\\
\Theta&=1-\frac{(\gamma-1)\epsilon^2}{4\alpha\gamma}G^2,
& \rho_0&=\Theta^{1/(\gamma-1)},\qquad P_0=\rho_0^\gamma.
\end{align*}
The nominal, weak and zero strengths are $\epsilon=0.3$, $0.03$ and $0$. For $x\ge1/2$, the state is uniform:
\[
\rho_2=\frac{(\gamma+1)M_1^2}{(\gamma-1)M_1^2+2},\qquad
P_2=1+\frac{2\gamma}{\gamma+1}(M_1^2-1),\qquad
u_2=\frac{U_\infty}{\rho_2},\qquad v_2=0.
\]
The left boundary prescribes the undisturbed upstream state $(\rho,P,u,v)=(1,1,U_\infty,0)$; the right boundary is outflow and the $y$ boundaries are reflecting. The SV marker fraction is zero. For both problems, $E_0=P_0/(\gamma-1)+\rho_0(u_0^2+v_0^2)/2$. Conserved cell averages are initialized with six-point Gauss--Legendre quadrature in each coordinate (36 points per cell). In the adaptive hierarchy, coarse initial states are averages of their four children. The SV shock lies on a cell face.

\subsection{Additional physical-field comparisons}

\begin{figure}[!ht]
\centering
\includegraphics[width=.84\linewidth]{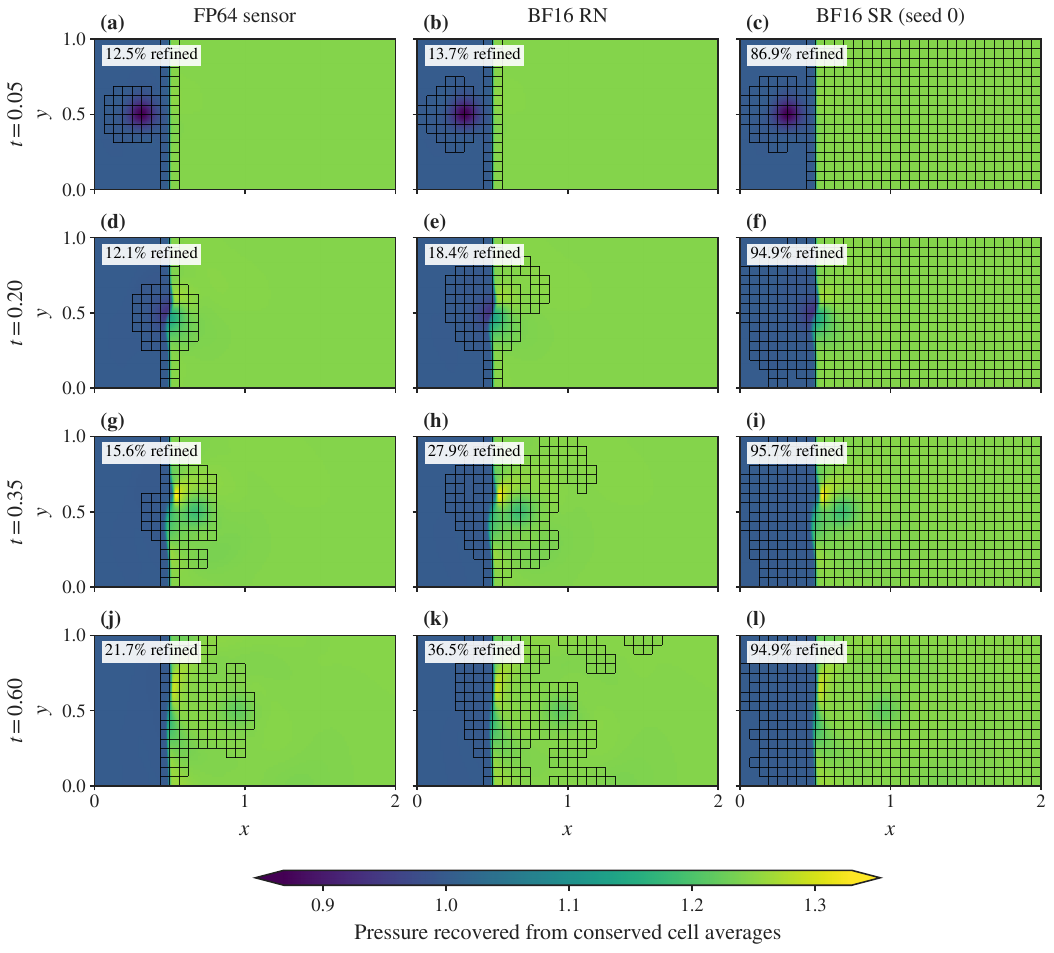}
\caption{Additional pressure snapshots and refinement blocks for shock--vortex interaction. The reference, BF16 RN and BF16 SR (seed zero) sensor policies are shown at common observation times.}
\label{fig:sv-history}
\end{figure}

Figure~\ref{fig:sv-history} shows four times of the SV comparison. BF16 SR selects an extended mesh in the downstream region throughout the interaction.
The short-time RT comparison at $32\times128$ gives mean refined areas of 12.50\% for BF16 RN and 12.52\% for BF16 SR, and mixing widths of 0.034032 and 0.034033, respectively, at $t=1$. In an initial comparison, 31 missed SR tags still produce the same 512 refined parents after buffering. The coarse adaptive and same-finest uniform FP64 references remain below mode amplitude 0.01 through $t=1$; the twice-finest uniform reference crosses between 0.875 and 1. To observe the nonlinear instability, the nominal problem is evolved to $t=3.75$ at base resolutions $32\times128$, $64\times256$ and $128\times512$. The HLL scheme, physical inputs and regrid interval $1/8$ are unchanged. With finest cell width $h_f=h/2$, each regrid interval uses $K=\lceil (1/8)/(0.4h_f/5)\rceil$ equal steps, where 5 is the prescribed bound on $|u|+|v|+2c$, with sound speed $c=\sqrt{\gamma P/\rho}$. This gives $K=400$, 800 and 1600 at the three base resolutions. The three policies are FP64 tagging, BF16 RN and BF16 independent SR. The observation at $t=3.75$ uses the active pre-remap geometry.

\begin{figure}[!ht]
\centering
\includegraphics[width=0.84\linewidth]{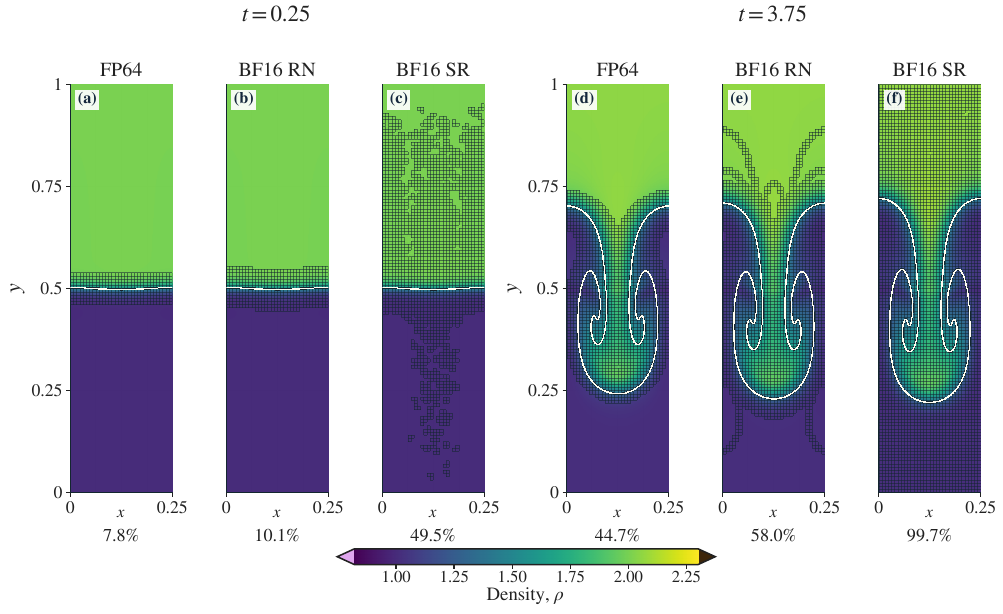}
\caption{\textbf{Development of the Rayleigh--Taylor instability on a $128\times512$ base mesh.} Native-density tagging with FP64, BF16 RN and BF16 SR (seed zero) at $t=0.25$ (a--c) and $t=3.75$ (d--f). Colors show cell-average density; dark lines mark refinement blocks before remapping and the white line marks the $C=0.5$ interface. Percentages denote the refined fraction of the domain.}
\label{fig:rt}
\end{figure}

The finest-grid late-time density and $C=0.5$ marker interface in Fig.~\ref{fig:rt} exhibit a descending heavy-fluid spike and roll-up. At $t=3.75$, 144 of 256 native fine-lattice columns have interface overhangs (more than one interface crossing) in the FP64 case, compared with 146 for each BF16 case. The lowest interface location changes from 0.24183 (FP64) to 0.22974 (RN) and 0.22061 (SR), and the highest from 0.70226 to 0.70952 and 0.72115. The mixing width $W_C=\int_0^1 4\overline C(y)[1-\overline C(y)]\,dy$, with $\overline C$ the horizontal mean marker fraction evaluated after conservative projection onto the common $128\times512$ grid, is 0.42602, 0.44297 and 0.46199 for the three policies. Their cumulative leaf-cell updates are $4.929\times10^9$, $5.756\times10^9$ and $9.931\times10^9$. The FP64 result on the intermediate base mesh gives $W_C=0.18083$; the corresponding RN and SR values are 0.18863 and 0.21137, with 20.90\%, 20.90\% and 56.74\% of the domain refined at the late observation. The lowest-resolution runs give identical fields and cumulative work ($6.7584\times10^7$ updates). The resolution dependence of the refinement threshold and the interpretation of the three-grid comparison are given in Section~\ref{sec:rt-results}. 

\subsection{State storage}
The state-storage controls separately convert initial states, states after each timestep, both, or reported-output arrays; another control combines state and native-sensor conversion. Physical admissibility requires $0\le m_C=\rho C\le\rho$, with positive density and pressure. If $m_C$ and $\rho$ lie in the same rounding interval at fractional positions $\theta_C\le\theta_\rho$, independent SR violates their ordering with probability $\theta_C(1-\theta_\rho)$. The exact rational marker test fails at the first conversion in 120 RT representation--seed cases (40 per format, five roles and eight seeds), including FP32; these include output-only violations without an invalid internal trajectory. Using a common variate would make the quantization map monotone and preserve this ordering.

\FloatBarrier

\section{Precision Atlas scope}\label{app:inventory}
Table~\ref{tab:inventory} lists the 60 kernels evaluated in the Atlas. Inputs, references and per-format results are given in the supplementary data. For 34 of the 120 scenarios the reference is the same discrete algorithm evaluated in FP64; these comparisons isolate arithmetic differences. 


\begin{table}[H]
\centering\footnotesize
\caption{The 60 kernels evaluated in the Precision Atlas.}
\label{tab:inventory}
\begin{tabularx}{\linewidth}{@{}l X l X@{}}
\toprule
ID & Algorithm & ID & Algorithm\\
\midrule
C01 & Rosenbrock--Euler Robertson & P05 & D1Q3 BGK collision\\
C02 & Arrhenius and net reaction rate & Q01 & Split-step Schr\"odinger FFT\\
C03 & Species normalization & Q02 & Two-site Hartree-like SCF\\
E01 & 1D Yee Maxwell & Q03 & Free 1D Wilson-type Dirac\\
E02 & Scalar Helmholtz LU & R01 & Absorption slab tally\\
E03 & Ideal-MHD physical flux & S01 & Field storage/casts\\
E04 & Cloud-in-cell deposit/gather & S02 & Stored time addition\\
E05 & Boris rotation & S03 & AXPY/state addition\\
F01 & FV RK2 advection & S04 & Sequential sums/dot products\\
F02 & FV RK2 diffusion & S05 & Small sequential GEMM\\
F03 & FV RK2 advection--diffusion & S06 & Periodic row action\\
F04 & WENO-JS5 / MC reconstruction & S07 & True residual\\
F05 & Rusanov/HLL/HLLC fluxes & S08 & Pivoted LU\\
F06 & 1D pressure projection & S09 & Weighted Jacobi\\
F07 & Conserved-to-primitive gas state & S10 & Unpreconditioned CG\\
F08 & Pressure-gradient extraction & S11 & Unpreconditioned GMRES\\
F09 & Local $k$--$\varepsilon$ balance & S12 & Block-Jacobi application\\
F10 & Graph curvature & S13 & Geometric multigrid V-cycles\\
F11 & D1Q3 BGK trajectory & S14 & Scalar constitutive Newton\\
F12 & Hydrostatic Rusanov balance & S15 & Finite-difference/analytic Jv\\
M01 & 1D elastic bar & S16 & Radix-2 FFT\\
M02 & Elastoplastic return map & S17 & Modified Gram--Schmidt QR\\
M03 & Penalty contact/Coulomb & S18 & Nodal linear prolongation\\
M04 & Elastic wave, velocity-Verlet & S19 & Triangular affine map\\
M05 & Two-field partitioned fixed point & S20 & Determinant-sign predicate\\
O01 & Discrete adjoint gradient & T01 & Implicit nonlinear conduction\\
P01 & All-pairs gravity & T02 & Enthalpy inversion\\
P02 & Lennard--Jones force/virial & T03 & Absorption/source sweep\\
P03 & Particle-mesh electrostatics & T04 & Blackbody exchange\\
P04 & One-bond SHAKE/RATTLE & T05 & Four-resistance heat network\\
\bottomrule
\end{tabularx}
\end{table}
\FloatBarrier

\Needspace{7\baselineskip}
\begingroup
\small\linespread{0.94}\selectfont
\setlength{\bibsep}{0pt}
\setlength{\parskip}{0pt}
\bibliographystyle{unsrtnat}
\bibliography{references}
\endgroup
\end{document}